\documentclass[twocolumn]{aastex63}
\usepackage{amsmath}
\usepackage{graphicx}
\usepackage{appendix}
\usepackage{comment}
\usepackage{tabularx}
\usepackage{booktabs}
\usepackage{comment}
\usepackage{xcolor}
\usepackage{hyperref}
\usepackage{lineno}

\newcommand{\h}{\mathbf{h}}
\newcommand{\Msun}{\ensuremath{M_\odot}}        
\newcommand{\Zsun}{\ensuremath{{\rm Z}_\odot}}  
\newcommand{\Lsun}{\ensuremath{L_\odot}}        
\newcommand{\Rsun}{\ensuremath{R_\odot}}        
\newcommand{\Mzams}{\ensuremath{M_{\rm ZAMS}}}  
\newcommand{\Mhe}{\ensuremath{M_{\rm He}}}      
\newcommand{\Teff}{\ensuremath{T_{\rm eff}}}    
\newcommand{\Tc}{\ensuremath{T_{\mathrm{c}}}}   
\newcommand{\rhoc}{\ensuremath{\rho_{\mathrm{c}}}}  
\newcommand{\Lnu}{\ensuremath{L_\nu}}           
\newcommand{\Lgamma}{\ensuremath{L_\gamma}}     
\newcommand{\LHe}{\ensuremath{L_{\rm He}}}      
\newcommand{\Enu}{\ensuremath{E_\nu}}           
\newcommand{\Egamma}{\ensuremath{E_\gamma}}     

\newcommand{\code}[1]{\texttt{#1}}
\newcommand{\mesa}{\code{MESA}}
\newcommand{\MESA}{\mesa}

\newlength{\apjcolwidth}
\begin{document}

\title{Stellar Neutrino Emission Across The Mass-Metallicity Plane}

\shorttitle{Stellar Neutrino Emission Across The Mass-Metallicity Plane}
\shortauthors{Farag et al.}

\author[0000-0002-5794-4286]{Ebraheem Farag}
\affiliation{School of Earth and Space Exploration, Arizona State University, Tempe, AZ 85287, USA}

\author[0000-0002-0474-159X]{F.X.~Timmes}
\affiliation{School of Earth and Space Exploration, Arizona State University, Tempe, AZ 85287, USA}

\author[0000-0002-5107-8639]{Morgan T. Chidester}
\affiliation{School of Earth and Space Exploration, Arizona State University, Tempe, AZ 85287, USA}

\author[0000-0002-5490-2689]{Samalka Anandagoda}
\affiliation{Department of Physics and Astronomy, Clemson University, Kinard Lab of Physics, Clemson, SC 29634-0978, USA} 

\author[0000-0002-8028-0991]{Dieter H. Hartmann}
\affiliation{Department of Physics and Astronomy, Clemson University, Kinard Lab of Physics, Clemson, SC 29634-0978, USA}

\correspondingauthor{Ebraheem Farag}
\email{ekfarag@asu.edu}

\begin{abstract}
We explore neutrino emission from nonrotating, single star models across six initial metallicities and seventy initial masses from the zero-age main sequence to the final fate. Overall, across the mass spectrum, we find metal-poor stellar models tend to have denser, hotter and more massive cores with lower envelope opacities, larger surface luminosities, and larger effective temperatures than their metal-rich counterparts. Across the mass-metallicity plane we identify the sequence (initial CNO $\rightarrow$ $^{14}$N $\rightarrow$ $^{22}$Ne $\rightarrow$ $^{25}$Mg $\rightarrow$ $^{26}$Al $\rightarrow$ $^{26}$Mg $\rightarrow$ $^{30}$P $\rightarrow$ $^{30}$Si) as making primary contributions to the neutrino luminosity at different phases of evolution. For the low-mass models we find neutrino emission from the nitrogen flash and thermal pulse phases of evolution depend strongly on the initial metallicity. For the high-mass models, neutrino emission at He-core ignition and He-shell burning depends strongly on the initial metallicity. Anti-neutrino emission during C, Ne, and O burning shows a strong metallicity dependence with $^{22}$Ne($\alpha$,$n$)$^{25}$Mg providing much of the neutron excess available for inverse-$\beta$ decays. We integrate the stellar tracks over an initial mass function and time to investigate the neutrino emission from a simple stellar population. We find average neutrino emission from simple stellar populations to be 0.5--1.2 MeV electron neutrinos. Lower metallicity stellar populations produce slightly larger neutrino luminosities and average $\beta$ decay energies. This study can provide targets for neutrino detectors from individual stars and stellar populations. We provide convenient fitting formulae and open access to the photon and neutrino tracks for more sophisticated population synthesis models.
\end{abstract}


\section{Introduction} \label{s.intro}

The next core-collapse (CC) supernova in the Milky Way or one of its satellite galaxies will be an opportunity to observe the explosion of a massive star across the electromagnetic, gravitational, and particle spectrums. For example, neutrinos with energies $\lesssim$\,10 MeV  have played a prominent role in stellar physics  \citep{hirata_1987_aa,hirata_1988_aa,bionta_1987_aa,alekseev_1987_aa, bahcall_1989_aa,borexino-collaboration_2014_aa,borexino-collaboration_2018_aa,borexino-collaboration_2020_aa} and particle physics \citep{bahcall_1989_aa, ahmad_2002_aa, ackermann_2022_aa}. Maps of $\ge$\,1 TeV neutrinos from the Galactic plane are consistent with a diffuse emission model of neutrinos whose analysis includes the supernova remnant and pulsar wind nebula outcome(s) of CC events \citep{IceCube_Collaboratio_2023_aa}.

Ongoing technological improvements in detector masses, energy resolution, and background abatement  will allow the global {\it SuperNova Early Warning System} network \citep{al-kharusi_2021_aa}  to observe new signals from different stages of the lifecycle of individual stars or the aggregate signal from a stellar population with multi-kiloton detectors such as {\it SuperKamiokande} \citep{simpson_2019_aa, harada_2023_aa}, {\it SNO+} \citep{allega_2023_aa}, {\it KamLAND} \citep{abe_2023_aa}, {\it Daya Bay} \citep{an_2023_aa}, {\it DUNE} \citep{Acciarri:2016ooe}, {\it JUNO} \citep{yang_2022_aa} and the upcoming {\it HyperKamiokande} \citep{abe_2016_aa}.

Examples of ongoing stellar neutrino searches include pre-supernova neutrinos which allow new tests of stellar and neutrino physics \citep[e.g.,][]{2004APh....21..303O, kutschera_2009_aa, odrzywolek_2009_aa, patton_2017_aa, patton_2017_ab, kato_2017_aa, kato_2020_aa, kosmas_2022_aa} and enable an early alert of an impending CC supernova to the electromagnetic and gravitational wave communities \citep{beacom_1999_aa, vogel_1999_aa, mukhopadhyay_2020_aa, al-kharusi_2021_aa}. Other ongoing explorations include the diffuse supernova neutrino background \citep{hartmann_1997_aa, 1984NYASA.422..319B, 1984Natur.310..191K,ando_2004_aa, horiuchi_2009_aa, beacom_2010_aa, anandagoda_2020_aa, suliga_2022_aa,anandagoda_2023_aa}, and neutrinos from the helium-core nitrogen flash \citep{serenelli_2005_aa}, compact object mergers \citep{2018PhRvD..97j3001K, 2020PhRvD.101b3016L}, tidal disruption of stars \citep{lunardini_2017_aa, winter_2022_aa, reusch_2022_aa}, and pulsational pair-instability supernovae \citep{leung_2020_aa}.

\cite{farag_2020_aa} introduced the idea of a neutrino Hertzsprung–Russell Diagram ($\nu$HRD) with a sparse grid of models. Each model started from the zero-age main sequence (ZAMS) and ended at a final fate but only at solar metallicity. They found all masses produce a roughly constant neutrino luminosity \Lnu \ during core H burning on the main-sequence (MS), and confirmed that low-mass (\Mzams\,$<$ 8 \Msun) Red Giant Branch (RGB) models with \Mzams\,$\leq$ 2 \Msun \ undergo large increases in \Lnu \  during the helium flash \citep[nitrogen flash for neutrinos,][]{serenelli_2005_aa} and subsequent sub-flashes. They also found He burning in asymptotic giant branch (AGB) models undergo sharp increases in \Lnu \  from thermal pulses (TPs), and significantly larger \Lnu \ from the hotter and denser cores of later evolutionary stages culminating at the onset of CC in high-mass (\Mzams\,$\geq$\,8~\Msun), non-rotating, single star models. A photon Hertzsprung–Russell Diagram ($\gamma$HRD) provides information about the stellar surface, a $\nu$HRD can serve as a diagnostic tool of the stellar interior.

Changes in the initial metallicity Z of a model changes the structure of the model through the equation of state \citep[EOS][]{saumon_1995_aa, timmes_2000_ab, rogers_2002_aa, irwin_2004_aa, potekhin_2010_aa, jermyn_2021_aa, bauer_2023_aa}, radiative opacity \citep{iglesias_1993_aa, iglesias_1996_aa, ferguson_2005_aa, ferguson_2008_aa, poutanen_2017_aa}, conductive opacity \citep{cassisi_2007_aa, blouin_2020_aa}, nuclear energy generation rate \citep{angulo_1999_aa, cyburt_2010_aa, sallaska_2013_aa, deboer_2017_aa, farag_2022_aa}, gravitational sedimentation \citep{bauer_2020_aa}, and mass-loss by line driven winds \citep{sanyal_2017_aa, vandenberg_2022_aa}.

The coupling between these pieces of stellar physics and neutrino production from thermal \citep{itoh_1996_aa} and weak reaction processes \citep{fuller_1985_aa,oda_1994_aa, langanke_2000_aa, nabi_2021_aa} suggests that changes in Z can cause changes in a $\nu$HRD, and upon integration, the neutrino emission from a simple stellar population model.

This article is novel in exploring stellar neutrino emission across the mass-metallicity plane. This study can provide targets for neutrino detectors from individual stars and stellar populations.
Section~\ref{s.input_physics} describes the mass-metallicity grid and stellar physics,
\S\,\ref{s.overall} presents overall features and drivers across the mass-metallicity plane, 
\S\,\ref{s.lowmass} analyzes low-mass tracks,
\S\,\ref{s.highmass} details high-mass tracks,
\S\,\ref{s.total_production} explores neutrino emission from a simple stellar population model, and
\S\,\ref{s.conc} summarizes our results.

Important symbols are defined in Table \ref{table:list-of-symbols}.
Acronyms and terminology are defined in Table \ref{tab:acronym}.

\startlongtable
\begin{deluxetable}{clc}
  \tablecolumns{3}
  \tablewidth{1.0\apjcolwidth}
  \tablecaption{Important symbols.
   \label{table:list-of-symbols}}
  \tablehead{\colhead{Name} & \colhead{Description} & \colhead{Appears}}
  \startdata
$A$            & Atomic number                       & \ref{s.input_physics}  \\
$D$            & Element diffusion coefficient       & \ref{s.input_physics}  \\
$E$            & Energy                              & \ref{s.all_Metallicity}  \\
$\epsilon$     & Average neutrino energy             & \ref{s.honeZ}  \\ 
$g$            & Gravitational acceleration          & \ref{s.all_Metallicity}  \\
$H$            & Pressure scale height               & \ref{s.all_Metallicity}  \\
$\kappa$       & Opacity                             & \ref{s.all_Metallicity}  \\
$k_{\rm B}$    & Boltzmann constant                  & \ref{s.all_Metallicity}  \\
$L$            & Luminosity                          & \ref{s.intro}  \\
$M$            & Stellar mass                        & \ref{s.intro}  \\
$\mu$          & Mean molecular weight               & \ref{s.input_physics}  \\
$n$            & Number density                      & \ref{s.input_physics}  \\
$\eta$         & Neutron excess                      & \ref{s.input_physics}  \\
$R$            & Stellar Radius                      & \ref{s.lMetallicity}  \\
$\rho$         & Mass density                        & \ref{s.all_Metallicity}  \\
$P$            & Pressure                            & \ref{s.all_Metallicity}  \\
$T$            & Temperature                         & \ref{s.input_physics}  \\
$\tau$         & Time or Timescale                    & \ref{s.lMetallicity}  \\
$X$            & Mass fraction                       & \ref{s.input_physics}  \\
X              & Hydrogen mass fraction              & \ref{s.input_physics}  \\
$Y$            & Abundance                           & \ref{s.input_physics}  \\
Y              & Helium mass fraction                & \ref{s.input_physics}  \\
$Y_e$          & Electron fraction                   & \ref{s.input_physics} \\
Z              & Metal mass fraction                 & \ref{s.intro}  \\
$Z$            & Atomic charge                       & \ref{s.input_physics}  \\
  \enddata
\tablenotetext{}
{\hsize \apjcolwidth {\bf Note:} 
Some symbols may be further subscripted, for example,
by $\mathrm{c}$ (indicating a central quantity),
by $\mathrm{\gamma}$ (indicating a photon quantity), 
or by $\mathrm{\nu}$ (indicating a neutrino quantity).
}
\end{deluxetable}

\startlongtable
\begin{deluxetable}{>{\raggedleft\arraybackslash}p{0.18\apjcolwidth} >{\centering\arraybackslash} p{0.65\apjcolwidth} >{\raggedright\arraybackslash}p{-0.10\apjcolwidth}}
  \tablecolumns{3}
  \tablewidth{1.0\apjcolwidth}
  \tablecaption{Acronyms and Terminology.\label{tab:acronym}}
  \tablehead{\colhead{Acronym} & \colhead{Description} & \colhead{Appears}}
  \startdata
AGB         & Asymptotic Giant Branch        & \ref{s.intro}  \\
CC          & Core Collapse                  & \ref{s.intro}  \\
CHeB        & Core Helium Burning            & \ref{s.overall}  \\
CHeD        & Core Helium Depletion          & \ref{s.overall}  \\
CO          & Carbon-Oxygen                  & \ref{s.all_oneZ} \\
EOS         & Equation of State              & \ref{s.intro}  \\
$\gamma$HRD & Photon Hertzsprung Russell Diagram    & \ref{s.intro}  \\
$\nu$HRD    & Neutrino Hertzsprung Russell Diagram    & \ref{s.intro}  \\
HB          & Horizontal Branch              & \ref{s.lowmass}  \\
IMF         & Initial mass function          & \ref{s.total_production} \\
MLT         & Mixing Length Theory           & \ref{s.input_physics}  \\
PMS         & Pre-Main Sequence              & \ref{s.input_physics}  \\
RGB         & Red Giant Branch               & \ref{s.intro}  \\
RSG         & Red Supergiant                 & \ref{s.highmass} \\
TAMS        & Terminal Age Main Sequence     & \ref{s.all_oneZ}  \\
TP          & Thermal Pulse                  & \ref{s.intro}  \\
WD          & White Dwarf                    & \ref{s.input_physics}  \\
ZAMS        & Zero Age Main Sequence         & \ref{s.intro}  \\
TAMS        & Terminal Age Main Sequence     & \ref{s.overall}  \\
Low-mass    & \Mzams\,$<$ 8 \Msun            & \ref{s.intro}  \\
High-mass   & \Mzams\,$\ge$ 8 \Msun          & \ref{s.intro}  \\
  \enddata
\end{deluxetable}

\begin{figure*}[!htb] 
    \centering
    \includegraphics[width=7.1in]{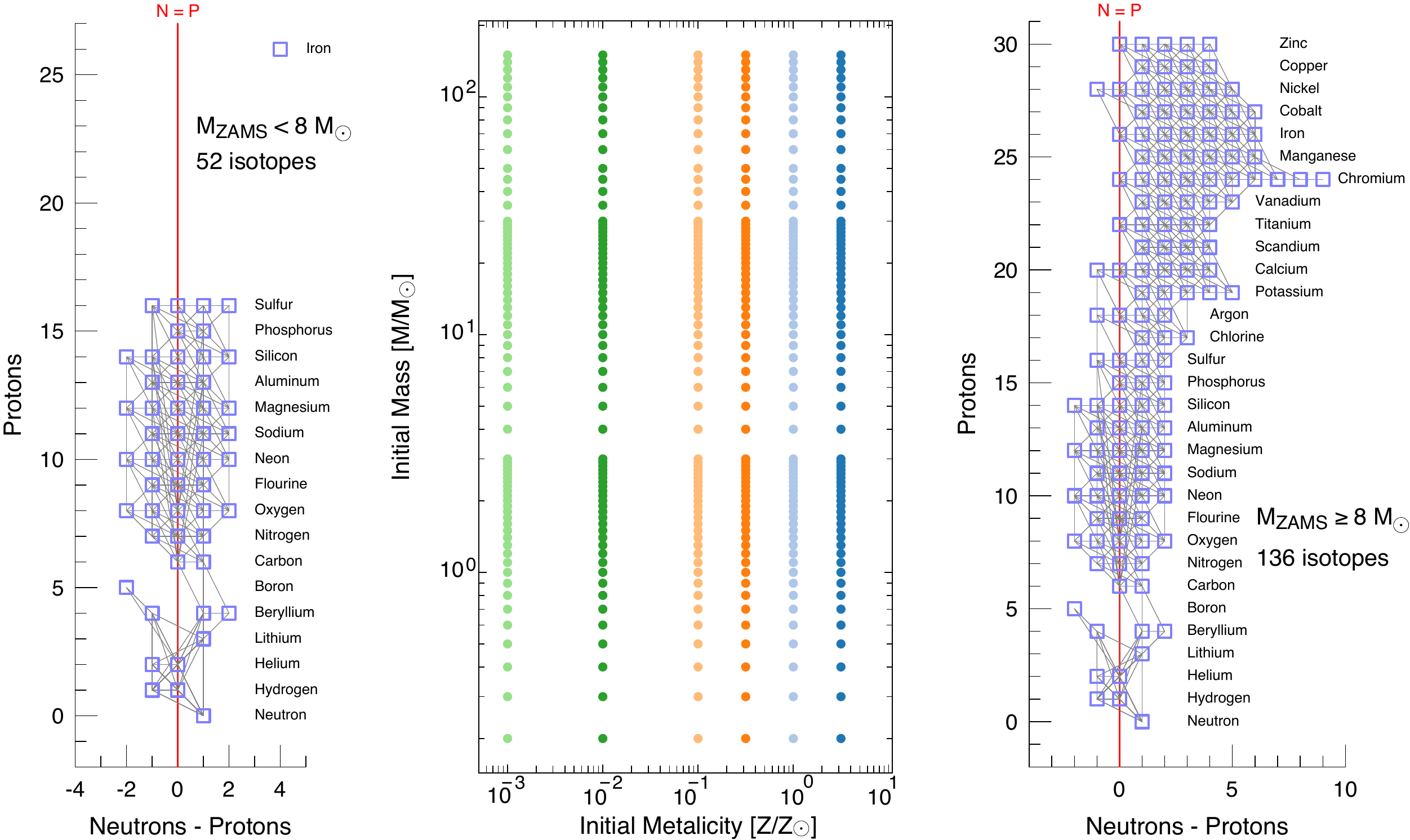}
    \caption{Coverage in the mass-metallicity plane (center). The x-axis is the initial Z of a model relative to solar, and the y-axis is \Mzams \ of a model relative to solar. Six metallicities, each marked with a different color, and 70 masses at each metallicity (circles) span the mass-metallicity plane. The nuclear reaction network for low-mass (left) and high-mass (right) models is illustrated. These x-axes are the difference in the number of neutrons and protons in an isotope. Positive values indicate neutron-rich isotopes, the zero value is marked by the red vertical line, and negative values indicate proton-rich isotopes. These y-axes are the number of protons in an isotope, labelled by their chemical element names. Isotopes in the reaction network are shown by purple squares. Reactions between isotopes are shown by gray lines. Note Fe in the low-mass reaction network does not react with other isotopes. Fe is included for a more consistent specification of the initial composition, hence any microphysics that depends upon the composition including the opacity, equation of state, element difffusion, and neutrino emission.}
    \label{fig:mznet}
\end{figure*}

\begin{figure*}[!htb]
    \centering
    \includegraphics[width=7.1in]{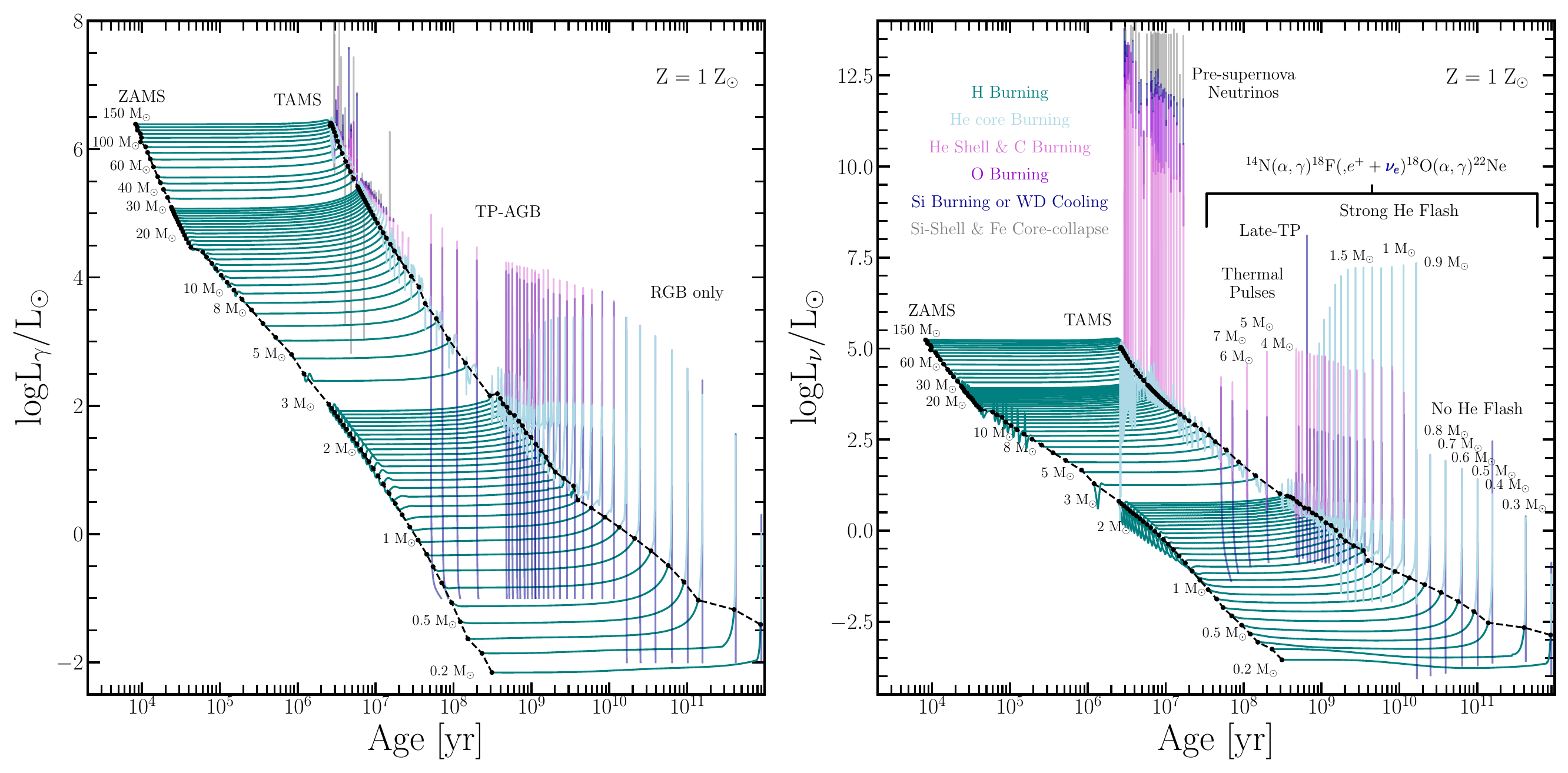}
    \caption{Light curves for photons (left) and neutrinos (right). Tracks span 0.2--150 \Msun \  for Z\,=\,1\,\Zsun\ and are labeled. Key phases of evolution including the ZAMS (black circles), TAMS (black circles), core He flashes (light green), thermal pulses, and pre-supernova stage are also labeled. The PMS light curves are suppressed for visual clarity. \Lnu \ during the nitrogen flash (He flash for photons) and thermal pulses for the M\,$<$\,8\,\Msun \ light curves can exceed \Lgamma.  At and beyond core C-burning  \Lnu \ dominates the evolution of the M\,$\ge$\,8\,\Msun \  light curves. Luminosities are normalized to $\Lsun = 3.828 \times 10^{33}$ erg s$^{-1}$ \citep{prsa_2016_aa}.}
    \label{fig:age_vs_luminosity_1zun}
\end{figure*}

\section{Mass-Metallicity Plane and Stellar Physics}\label{s.input_physics}

We model the evolution of single, non-rotating stars over a wide range of initial masses and metallicities, from the pre-main sequence (PMS) to the final fate. Figure~\ref{fig:mznet} shows the mass-metallicity plane for 70 \Mzams \ models distributed in the range 0.2\,\Msun\, $\le$ \Mzams $\le$ 150\;\Msun\ for six initial metallicities $\log$(Z/\Zsun)\,=\,0.5, 0, $-$0.5, $-$1, $-$2, $-$3, where we choose \Zsun = 0.0142 \citep{asplund_2009_aa}. This mass-metallicity plane spans the range of single stars found in the Galaxy \citep{edvardsson_1993_aa, ratcliffe_2023_aa, almeida-fernandes_2023_aa}, and aids estimates of the neutrino emission from simple stellar population models.

We use \MESA\ version r15140 to construct our stellar models \citep{paxton_2011_aa,paxton_2013_aa,paxton_2015_aa,paxton_2018_aa,paxton_2019_aa,jermyn_2023_aa}. We follow MIST \cite[][]{choi_2016_aa} to scale the H mass fraction X, He mass fraction Y, and metallicity Z
\begin{eqnarray}
{\rm Y} &=& {\rm Y}_{\rm p} + \left (\frac{{\rm Y}_{\odot}-{\rm Y}_{\rm p}}{{\rm Z}_{\odot}} \right) {\rm Z} \\ 
{\rm X} &=& 1-{\rm Y}-{\rm Z} \;\;,
\end{eqnarray}
where we adopt the primordial He abundance Y$_{\rm p}$\,=\,0.249 \citep{planck-collaboration_2016_aa}, $Y_{\odot}$\,=\,0.2703 and $Z_{\odot}$\,=\,0.0142 with mass fractions from \cite{asplund_2009_aa}.

For the low-mass models, we chose the Riemers wind mass loss scheme \citep{reimers_1977_aa} with an efficiency of 0.5 on the RGB, and Bl\"{o}ckers wind mass loss scheme \citep{blocker_2001_aa} with an efficiency of 1.0 on the AGB. All low-mass models terminate as a white dwarf (WD) at $L$\,=\,10$^{-3}$\,\Lsun, even if the evolution is longer than the age of the universe.  

For the high-mass models, we choose the ``Dutch'' wind loss scheme \citep{nieuwenhuijzen_1990_aa, nugis_2000_aa,vink_2001_ab,glebbeek_2009_aa} with an efficiency of 1.0 to generate stripped models. All models use an Eddington-grey iterated atmosphere as an outer boundary condition. We apply an extra pressure to the surface \citep[see Section 6.1 of][]{jermyn_2023_aa} of our AGB and high-mass models to maintain stability of the surface layer in super Eddington regimes where the surface of the model can otherwise run away. The termination age for all high-mass models is at the onset of CC when the infall velocity of the Fe core reaches 100\,km~s$^{-1}$. A subset of our models halted prematurely: at core C-depletion (\Mzams\,=\,8--10~\Msun), a stalled Ne/O flame in a degenerate core (\Mzams\,=\,11--14~\Msun), the onset of pair-instability (C-ignition with M\textsubscript{He}\,$\gtrsim$\,45~\Msun), or due to numerical difficulties near the onset of CC.

We adopt a minimum chemical diffusive mixing coefficient of D$_{\rm{min}}$\,=\,$10^{-2}$ cm\textsuperscript{2} s$^{-1}$ from C-ignition to the onset of CC to aid the convergence properties of our high-mass models \citep{farag_2022_aa}. To reduce the numerical cost we use operator splitting to decouple the hydrodynamics from the nuclear burning for temperatures above $T$\,=\,1$\times$10\textsuperscript{9}~K \citep{jermyn_2023_aa}.

We also adopt $\alpha$\,=\,1.5 for the convective mixing-length parameter, and $f_{ov}$\,=\,0.016, $f_{0,ov}$\,=\,0.008 for the convective overshooting parameters in all convective regions \citep{herwig_2000_aa, choi_2016_aa}. All stellar models use the MLT++ treatment for superadiabatic convection in the envelopes \citep{Sabhahit_2021_aa}. We also damp the velocities in the envelopes of our low-mass AGB models and high-mass models during the advanced burning stages to inhibit the growth of radial pulsations.

Figure \ref{fig:mznet} illustrates the 52 isotope nuclear reaction network used for low-mass stars and the 136 isotope reaction network used for high-mass models. Extended networks are required to accurately capture the nuclear energy generation, composition and stellar structure profiles, and the neutrino luminosity and spectra from $\beta$-processes \citep{farmer_2016_aa,patton_2017_aa,patton_2017_ab,kato_2020_aa}. The 136 isotope network  is reliable up to the onset of Si-shell burning, $T$\,$\lesssim$\, 4$\times$10$^{9}$~K. At higher temperatures, the paucity of Fe group isotopes in this reaction network cannot fully capture the nuclear burning \citep{farmer_2016_aa,patton_2017_aa}.

Nuclear reaction rates are a combination of NACRE \citep{angulo_1999_aa} and JINA REACLIB \citep{Cyburt_2010_ab}. We use the median \textsuperscript{12}C($\alpha$,$\gamma$)\textsuperscript{16}O  reaction rate from the experimental probability distribution function provided by \citet{deboer_2017_aa}, updated in \citet{mehta_2022_aa}, and publicly released in \citet{chidester_2022_aa}. Reaction rate screening corrections are from \citet{chugunov_2007_aa}, which includes a physical parameterization for the intermediate screening regime and reduces to the weak \citep{dewitt_1973_aa, graboske_1973_aa} and strong \citep{alastuey_1978_aa,itoh_1979_aa} screening limits at small and large values of the plasma coupling parameter. Weak reaction rates are based, in order of precedence, on  \citet{langanke_2000_aa}, \citet{oda_1994_aa}, and \citet{fuller_1985_aa}.

Baryon number is conserved in nuclear reactions. Define the abundance of species $Y_i$ by 
\begin{equation}
Y_i = \frac{n_i}{n_B} 
\end{equation}
where 
$n_i$ is the number density of isotope $i$ 
and $n_B$ is baryon number density. 
The number of baryons in isotope $i$ divided by the total number of baryons of all isotopes is 
the baryon fraction (mass fraction)
\begin{equation}
X_i = \frac{n_i \ A_i}{n_B}  = Y_i \ A_i 
\ ,
\end{equation}
where $A_i$ is the atomic mass number, the number of baryons in an isotope.
The mean atomic number is 
\begin{equation}
\overline{A} = \frac{\sum n_i { A}_i}{\sum n_i}  = \frac{n_B}{\sum n_i}  
= \frac{\sum Y_i {A}_i}{\sum Y_i}  = \frac{1}{\sum Y_i} 
\ ,
\end{equation}
the mean charge is
\begin{equation}
\overline{Z} = \frac{\sum n_i {Z}_i}{\sum n_i}  = \frac{\sum Y_i {Z}_i}{\sum Y_i}  = \overline{A} \sum Y_i {Z}_i
\ ,
\end{equation}
the electron to baryon ratio (electron fraction) is  
\begin{equation}
Y_e = \frac{n_e}{n_B} = \frac{\sum n_i Z_i}{n_B} = \sum Y_i Z_i = \frac{\overline{Z}}{\overline{A}}
\ ,
\end{equation}
where  $n_e$ is the free electron number density and the second equality assumes full ionization.
The related neutron excess is
\begin{equation}
\eta = \sum ({\rm N}_i - {\rm Z}_i) Y_i 
     = 1 - 2 Y_e 
     \ ,
\end{equation}
the mean ion molecular weight is
\begin{equation}
\mu_{{\rm ion}} = \overline{A}
\ ,
\end{equation}
the mean electron molecular weight is
\begin{equation}
\mu_{{\rm ele}} = \frac{1}{Y_e} = \frac{\overline{A}}{\overline{Z}}
\ ,
\end{equation}
and the mean molecular weight is 
\begin{equation}
\mu = \left [ \frac{1}{\mu_{\rm ion}} + \frac{1}{\mu_{\rm ele}} \right ]^{-1}
= \frac{\overline{A}}{\overline{Z} + 1}
= \frac{ n_B}{\sum n_i + n_e}
\ .
\end{equation}

Across the mass-metallicity plane the dominant thermal neutrino processes in our models are plasmon decay ($\gamma_{\rm plasmon} \rightarrow \nu + \bar{\nu}$) which scales with the composition as $Y_e^3$, 
photoneutrino production ($e^- + \gamma \rightarrow e^- + \nu + \bar{\nu}$) which scales as $Y_e$, and pair annihilation ($e^- + e^+ \rightarrow \nu + \bar{\nu}$) which also scales as $Y_e$.
All else being equal, as material becomes more neutron rich the neutrino emission from these three dominant processes decrease. 

Bremsstrahlung ($e^- + ^A$\!$Z \rightarrow e^- + ^A$\!$Z + \nu + \bar{\nu}$), which scales with the composition as $Y_e \overline{Z}$, and recombination ($e^-_{\rm continuum} \rightarrow e^-_{\rm bound} + \nu + \bar{\nu}$) , which scales as $\overline{Z}^{14}/\overline{A}$, play smaller roles. Neutrino emission from these five processes are discussed in \citet{itoh_1989_aa,itoh_1992_aa,itoh_1996_aa,itoh_1996_ab,kantor_2007_aa} and implemented, with partial first derivatives, in the \MESA\ module \code{neu}. Differential thermal neutrino emission rates are discussed in \cite{ratkovic_2003_aa,dutta_2004_aa,misiaszek_2006_aa,odrzywoek_2007_aa,kato_2015_aa,patton_2017_aa,dzhioev_2023_aa}.

Each of the 420 stellar models in the mass-metallicity grid use between 2000--3500 mass zones (lower values occur at ZAMS where there are no composition gradients) with $\simeq$3000 mass zones over the evolution being typical. 
Each low-mass model uses 1$\times$10$^{5}$ -- 3$\times$10$^{5}$ timesteps depending on the number of thermal pulses (TPs), and each high-mass model uses 2$\times$10$^{4}$ -- 5$\times$10$^{4}$ timesteps.
Each model executes on a 16 core node with 2~GHz AMD Epyc 7713 CPUs,  with low-mass models consuming 14--21 days and high-mass models using 10-21 days. The uncompressed total data set size is $\simeq$\,730 GB. 

The \MESA\ files to reproduce our models, and open access to the photon and neutrino tracks, are available at 
\dataset[http://doi.org/10.5281/zenodo.8327401]{http://doi.org/10.5281/zenodo.8327401}. 

\section{Overall Mass-Metallicity Features}\label{s.overall}

Here we present features and drivers of the neutrino emission, first at one metallicity in Section~\ref{s.all_oneZ} and then for all six metallicities in Section~\ref{s.all_Metallicity}.

\subsection{One Metallicity}\label{s.all_oneZ}

Figure~\ref{fig:age_vs_luminosity_1zun} shows the photon and neutrino light curves for all 70 calculated masses at Z\,=\,1\,\Zsun. Both plots begin at the ZAMS, defined when the luminosity from nuclear reactions $L_{\rm nuc}$ is 99\% of the total luminosity $L$, marking a transition from evolution on thermal timescale to a nuclear timescale. 

MS evolution is characterized by stable core H-burning, where neutrinos are produced by weak reactions in the proton-proton (pp) chains
p(p,e$^+$$\nu_e$)$^2$H, 
p(e$^-$p,$\nu_e$)$^2$H, 
$^3$He(p,e$^+$$\nu_e$)$^4$He, 
$^7$Be(e$^-$,$\nu_e$)$^7$Li, 
$^8$B(,e$^+$$\nu_e$)$^8$Be,
and the CNO cycles 
$^{13}$N(,e$^+$$\nu_e$)$^{13}$C, 
$^{13}$N(e$^{-}$,$\nu_e$)$^{13}$C,
$^{15}$O(,e$^+$$\nu_e$)$^{15}$N, 
$^{15}$O(e$^{-}$,$\nu_e$)$^{15}$N,
$^{17}$F(,e$^+$$\nu_e$)$^{17}$O, 
$^{17}$F(e$^{-}$,$\nu_e$)$^{17}$O,
$^{18}$F(,e$^+$$\nu_e$)$^{18}$O,
where electron capture reactions on CNO nuclei are included \citep{stonehill_2004_aa}. 

Models with \Mzams \, $\lesssim$ 1.2 \Msun \ have a central temperature $\Tc \lesssim$ 18$\times$10$^{7}$~K and burn H in their cores primarily through the pp chains, with a small fraction from the CNO cycles. For example, based on observations of solar neutrinos CNO burning accounts for around 1.6\% of the current energy generation of the Sun \citep{naumov_2011_aa,borexino-collaboration_2020_aa}. Models with \Mzams \, $\gtrsim$ 1.2 \Msun\ have $\Tc \gtrsim$ 18$\times$10$^{7}$~K and maintain their stability primarily from the CNO cycles \citep{wiescher_2010_aa}. Metal-poor models can produce their own carbon to begin CNO cycle H-burning  \citep{mitalas_1985_aa,wiescher_1989_aa,weiss_2000_aa,tompkins_2020_aa}. In addition, most of a model's initial Z comes from the CNO and $^{56}$Fe nuclei inherited from its ambient interstellar medium.  The slowest step in the CNO cycle is proton capture onto $^{14}$N, resulting in all the CNO catalysts accumulating into $^{14}$N during core H-burning.

All light curves in Figure~\ref{fig:age_vs_luminosity_1zun} proceed to the terminal age main sequence (TAMS), defined by core H-depletion (X$_{\rm c}$\,$\le$\,10$^{-6}$). The He-rich core contracts as a H-burning shell forms. The higher temperatures of shell H-burning can activate the Ne-Na, and Mg-Al cycles \citep{salpeter_1955_ab,marion_1957_aa,arnould_1999_aa,jose_1999_aa, izzard_2007_aa, boeltzig_2022_aa}. The light curves then bifurcate depending on \Mzams.

During He-burning the accumulated $^{14}$N is converted into the neutron-rich isotope $^{22}$Ne through the reaction sequence $^{14}$N($\alpha$,$\gamma$)$^{18}$F(,$e^{+}\nu_e$)$^{18}$O($\alpha$,$\gamma$)$^{22}$Ne, also shown in Figure~\ref{fig:age_vs_luminosity_1zun}.  This sequence is the source of neutrinos powering L$_{\nu}$ through all phases of He-burning \citep{serenelli_2005_aa, farag_2020_aa}.  
 
Usually the ashes of nuclear burning have a larger $\overline{A}$ and lie interior to the unburned fuel. For example, a He core is interior to a H-burning shell, and a carbon-oxygen (CO) core is interior to a He-burning shell. Exceptions occur when electron degeneracy and thermal neutrino losses lead to a temperature inversion with cooler temperatures in the central regions and hotter temperatures exterior to the core. The fuel ignites off-center and a burning front propagates towards the center.

For example, the 0.9\,\Msun\,$\le$\,\Mzams\,$\le$\,2\,\Msun \ light curves in Figure~\ref{fig:age_vs_luminosity_1zun} undergo off-center He ignition, the He Flash \citep{thomas_1967_aa,bildsten_2012_ab,gautschy_2012_aa,serenelli_2017_aa}. The accompanying nitrogen flash for neutrinos \citep{serenelli_2005_ab} are prominent and labeled. In contrast, the \Mzams\,$\ge$\,2\,\Msun\ light curves undergo central He burning. The 0.9\,\Msun\,$\le$\,\Mzams\,$\le$\,7\,\Msun \ light curves undergo TPs on the AGB, generating neutrinos first from H burning, and subsequently from He burning. A few light curves show a late TP during the transition to a cool WD. 

Neutrino emission from nuclear reactions dominate whenever H and He burn, otherwise neutrinos from thermal processes generally dominate \citep{farag_2020_aa}. For example, light curves for \Mzams \, $\ge$ 8~\Msun \ in Figure~\ref{fig:age_vs_luminosity_1zun} have the minimum mass for C ignition and those for \Mzams \, $\ge$ 10~\Msun \ have the minimum mass for Ne ignition \citep{becker_1979_aa,becker_1980_aa,garcia-berro_1997_aa,farmer_2015_aa,de-geronimo_2022_aa}. For these advanced burning stages  \Lnu \ in Figure~\ref{fig:age_vs_luminosity_1zun} become nearly vertical and greatly exceeds \Lgamma. Thermal neutrinos from pair-production dominate until the last few hours before CC when neutrinos from nuclear processes contribute \citep{2004APh....21..303O,odrzywolek_2010_aa,patton_2017_aa,patton_2017_ab,kato_2020_aa,kato_2020_ab}.

\begin{figure}[!htb]
    \centering
    \includegraphics[width=3.3in]{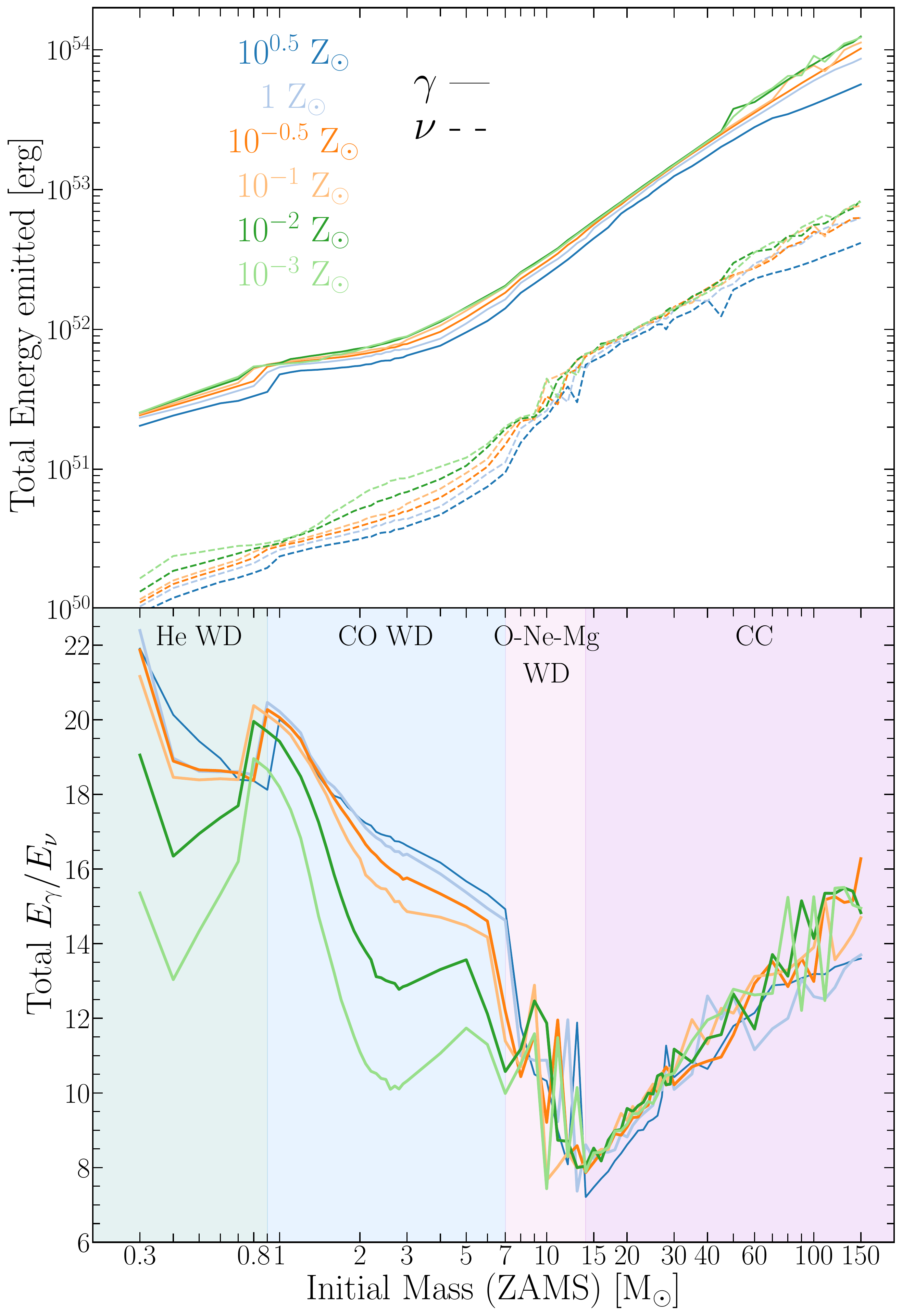}
    \caption{Total energy emitted in photons and neutrinos over the lifetime of a model (top) and their ratio (bottom) across the mass-metallicity plane. Transition between different final fates occur at local extrema, indicated by the colored panels and labels.}
    \label{fig:tot_energy_combined_mass}
\end{figure}

\subsection{Six Metallicities}\label{s.all_Metallicity}

The top panel of Figure~\ref{fig:tot_energy_combined_mass} shows the total energy emitted in photons $E_{\gamma}$ and neutrinos $E_{\nu}$, obtained by integrating \Lgamma \ and \Lnu \ over the lifetime of a model. Metal-poor models tend to have larger $E_{\gamma}$ and $E_{\nu}$ than the metal-rich models. 
Homology relations with power-law expressions for a bound-free Kramers opacity 
$\kappa \propto {\rm Z}(1 + {\rm X}) \rho T^{-3.5}$, 
pp-chain energy generation rate 
$\epsilon_{\rm pp} \propto {\rm X}^2 \rho T^4$, 
and mean molecular weight 
$\mu \propto {\rm X}^{-0.57}$ 
lead to  \citep{sandage_1986_aa,hansen_2004_ab}
\begin{equation}
    \Egamma \simeq \Lgamma \, \tau_{\rm MS} \propto (Z^{-1.1} X^{-5.0} M^{5.5}) \, \tau_{\rm MS} 
    \ ,
    \label{eq.lowmass}
\end{equation}
where $\tau_{\rm MS}$ is the MS lifetime.
Similarly, for a Thomson electron scattering opacity 
$\kappa \propto 1 + {\rm X}$
and CNO cycle energy generation rate 
$\epsilon_{\rm CNO} \propto {\rm X}  {\rm Z} \rho T^{17}$, 
\begin{equation}
    \Egamma \simeq \Lgamma \, \tau_{\rm MS} \propto (Z^{-1.0} X^{-4.3} M^{5.1}) \, \tau_{\rm MS} 
    \ .
    \label{eq.highmass}
\end{equation}

These expressions suggest that displacement on the MS due to a lower Z is partially offset by a shift to a larger X \citep{demarque_1960_aa}.  In addition, a lower Z requires higher \Tc \ to produce the same \Lgamma \ and \Lnu.  This is mainly why the low-Z high-mass models in Figure~\ref{fig:tot_energy_combined_mass} produce only a marginally larger \Lgamma \ and \Lnu \ on the MS while possessing larger \Tc. In turn, a larger \Lgamma \ implies a larger radiative gradient, and thus a larger core mass. 

\Lgamma \ and \Lnu \ in the core is primarily set by the mass of the model.
Envelope opacities affect the rate of nuclear reactions in the core insofar as the envelope has a large mass. The hotter the model is overall (e.g., the more massive), the less mass in the envelope will be cold enough to provide bound-free or bound-bound opacity. The largest differences due to the opacity occur in the low-mass models because they are colder, both in the core and the envelope. The models adjust the structure to accommodate a change in Z at a fixed luminosity.

Overall, across the mass spectrum, metal-poor stellar models tend to have denser, hotter and more massive cores with lower envelope opacities, larger surface luminosities and larger effective temperatures \Teff \ than their metal-rich counterparts \citep{demarque_1960_aa,iben_1963_aa,demarque_1967_aa,iben_1970_aa,vandenberg_1983_aa,sandage_1986_aa,hansen_2004_ab,georgy_2013_ab,young_2018_aa, groh_2019_aa,kemp_2022_aa}. These are the main drivers of changes to the thermal and nuclear reaction neutrino emission as the initial Z changes.

The bottom panel of Figure~\ref{fig:tot_energy_combined_mass} shows the ratio $E_{\gamma}$/$E_{\nu}$.
A maximum of $E_{\gamma}$/$E_{\nu}$\,$\simeq$\,20 at \Mzams\,$\simeq$\,0.9\,\Msun \ occurs at the transition between models that ignite He and those that do not, between the most massive He WD and the least massive CO WD. As \Mzams \ increases the resulting electron degenerate cores, first CO then ONeMg, become progressively more massive, denser, and hotter \citep[also see][]{woosley_2015_aa}. This increases production of thermal neutrinos from the plasmon, photoneutrino, and pair annihilation channels faster than the production of reaction neutrinos or photons. Thus $E_{\gamma}$/ $E_{\nu}$ decreases with \Mzams \ as shown in Figure~\ref{fig:tot_energy_combined_mass}. 

A minimum of $E_{\gamma}$/$E_{\nu}$\,$\simeq$\,8 at \Mzams\,$\simeq$\,12\,\Msun \ in Figure~\ref{fig:tot_energy_combined_mass} occurs at the transition between models that produce the most massive WD and those that go to CC. As \Mzams \ further increases, thermal neutrinos from pair annihilation increases slower than reaction neutrinos or photons, and thus $E_{\nu}$ is smaller than $E_{\gamma}$ in more massive models (pulsational pair-instability supernovae models are suppressed). The ratio $E_{\gamma}$/$E_{\nu}$ thus rises from the minimum and develops a roughly linear trend for \Mzams\,$\gtrsim$\,12\,\Msun. Overall, both extrema of $E_{\gamma}$/$E_{\nu}$ of Figure~\ref{fig:tot_energy_combined_mass} correlate with transitions in the final fate. 

\begin{figure*}[!htb]
    \centering
    \includegraphics[width=6.5in]{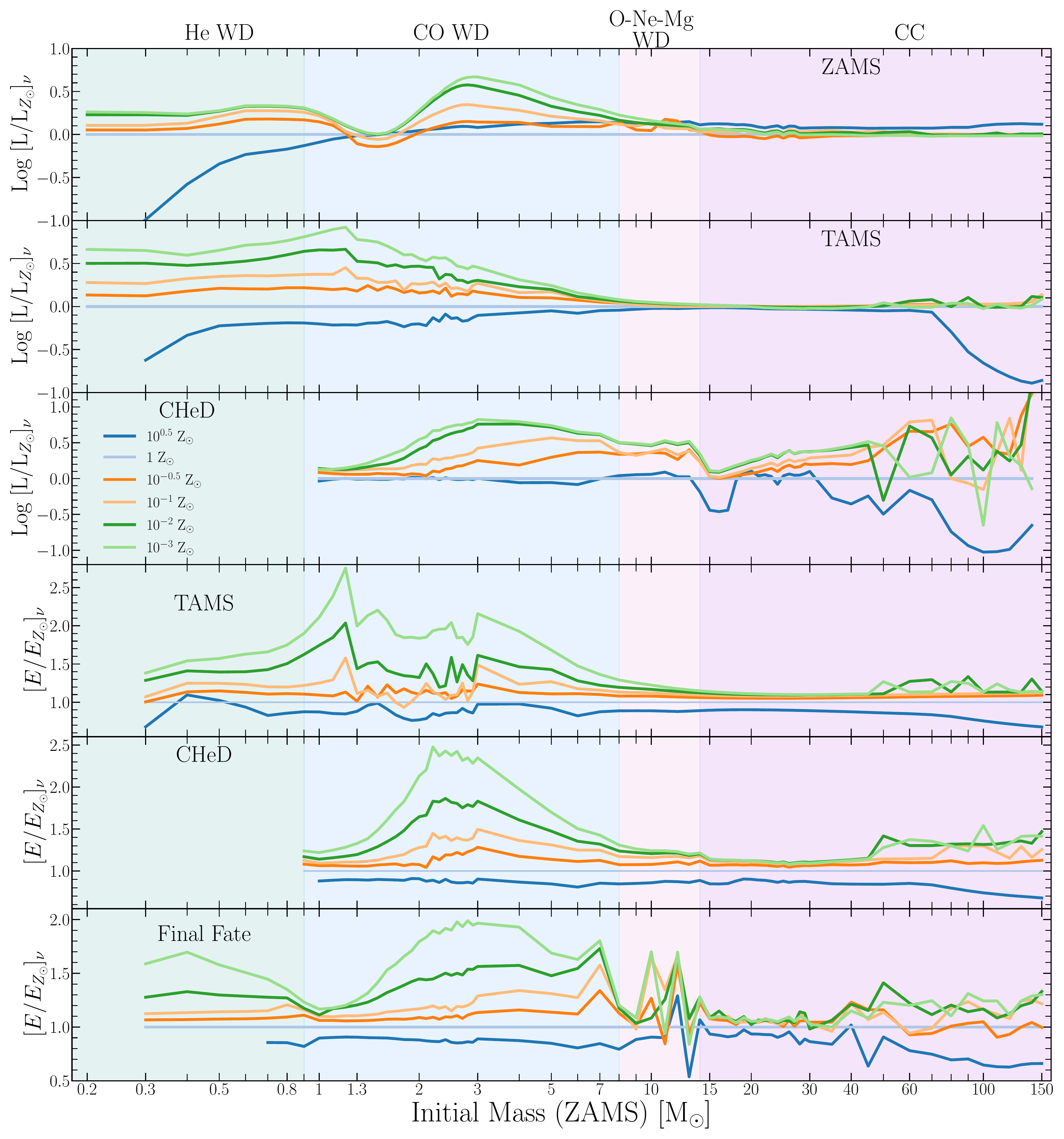}
    \caption{Ratio of \Lnu \ to \Lnu \ of the Z\,=\,1\,\Zsun \  model versus \Mzams \ for all six metallicities at ZAMS (top panel), TAMS (second panel) and CHeD (third panel).  The ratio of \Enu \ to \Enu \ of the Z\,=\,1\,\Zsun \ model versus \Mzams \  for all six metallicities at TAMS (fourth panel), CHeD (fifth panel) and final fate (bottom panel). Each panel is colored by the final fate given by the legend.}
    \label{fig:Lnu_ratio_ZAMS_and_TAMS}
\end{figure*}

\begin{figure*}[!htb] 
    \centering
    \includegraphics[width=7.1in]{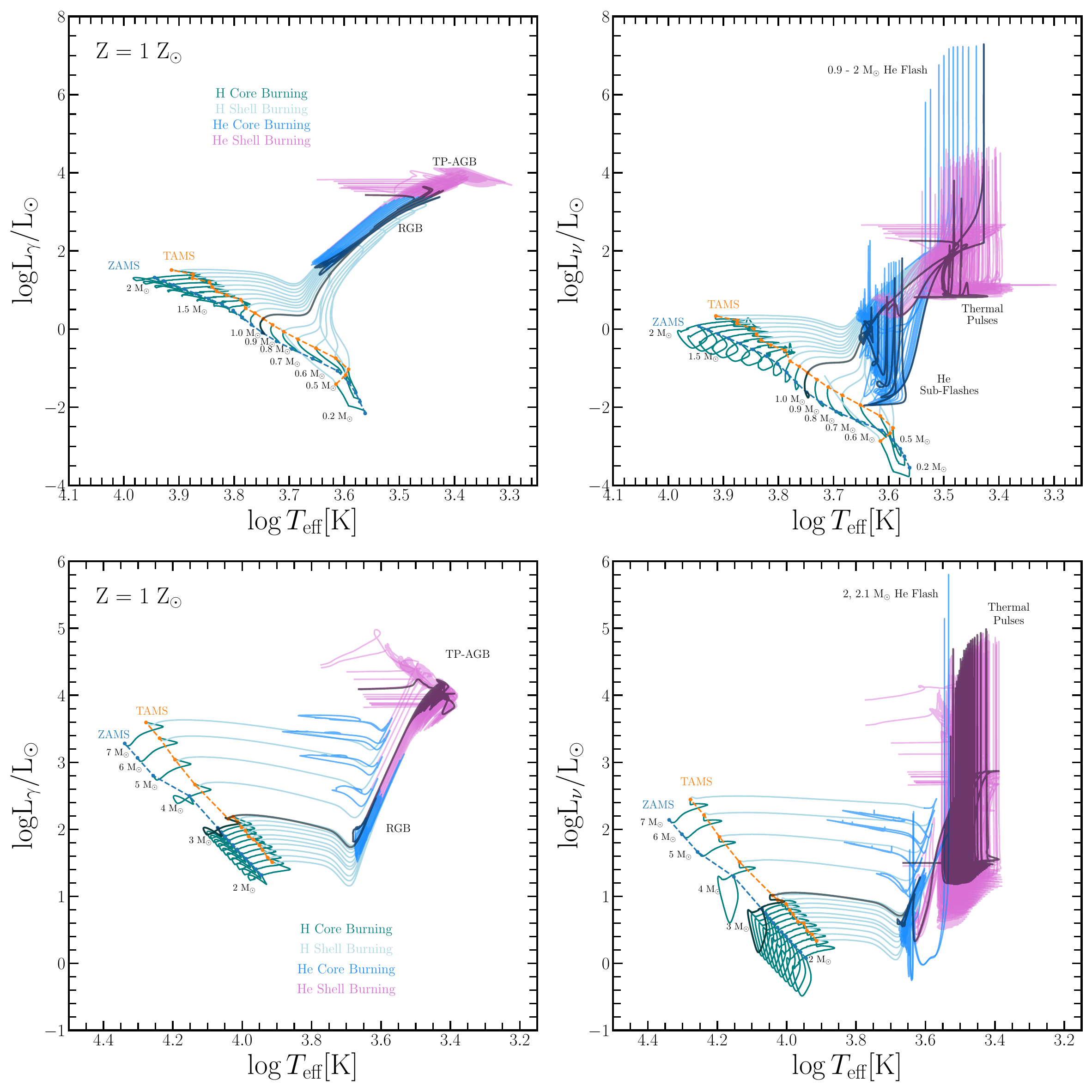}
    \caption{Low-mass tracks in a $\gamma$HRD (left panels) and a $\nu$HRD (right panels) for Z\,=\,1\,\Zsun \ over 0.2--2.0 \Msun\ (top row) and 2.0--7.0 \Msun (bottom row).  Tracks are colored by evolutionary phase and labeled. WD cooling tracks are suppressed for visual clarity.   Luminosities are normalized to $\Lsun = 3.828 \times 10^{33}$ erg s$^{-1}$ \citep{prsa_2016_aa}. The 1~\Msun\ and 3~\Msun\ tracks are highlighted in black as they are analyzed in detail as examples of low-mass models that do and do not undergo the He flash, respectively.}
    \label{fig:hr_lowmass}
\end{figure*}

Another trend in the bottom panel of Figure~\ref{fig:tot_energy_combined_mass} is the metallicity dependence of \Mzams \ models that become CO WD, the blue shaded region. More metal-rich models have a larger $E_{\gamma}$/$E_{\nu}$ than metal-poor models. A larger initial Z produces a larger accumulation of $^{14}$N during CNO cycle H-burning, thus a larger mass fraction of $^{22}$Ne during He-burning, and hence a smaller $Y_e$ as the CO WD becomes more neutron-rich. Plasmon neutrino rates scale as $Y_e^3$ leading to a smaller $E_{\nu}$, hence more metal-rich models have a larger $E_{\gamma}$/$E_{\nu}$ than metal-poor models in this \Mzams \ range. The dependence of CO WD on the $^{22}$Ne mass fraction, the degree of neutronization, may have implications for the progenitors Type Ia supernova \citep{timmes_2003_aa,townsley_2009_aa,bravo_2010_aa,piersanti_2022_aa} and the pulsation periods of variable WD \citep{campante_2016_aa,chidester_2021_aa,althaus_2022_aa}.

\cite{farag_2020_aa} showed \Lgamma/\Lnu\,$\simeq$\,40 for a standard solar model. As this model evolves off the MS the inert He core becomes denser, more electron degenerate, thermal neutrino production rise, \Lnu \ increases, and  thus \Lgamma/\Lnu decreases. Integrated over the lifetime of the model, $E_{\gamma}$/$E_{\nu}$ decreases to $\simeq$\,20 as shown in Figure~\ref{fig:tot_energy_combined_mass}. 

For any \Mzams, what is the impact of changing Z on the neutrino emission at any evolutionary stage?

Figure~\ref{fig:Lnu_ratio_ZAMS_and_TAMS} compares \Lnu \ to \Lnu \ of the Z\,=\,1\,\Zsun \ model across the mass-metallicity plane at three evolutionary stages in the top three panels.  As for Figure~\ref{fig:tot_energy_combined_mass}, at the ZAMS there is generally a small dependence on the initial Z but there are interesting features.  For example, the dip at \Mzams\,$\simeq$\,1.2\,\Msun \ corresponds to the transition from pp-chain dominated to CNO cycle dominated H-burning. Another feature is the stronger Z dependence for \Mzams \ models that become CO WD. As low-Z models tend to have denser, hotter and more massive H-burning cores, thermal and reaction neutrino contributions to \Lnu \ is larger relative to high-Z models.

At the TAMS, the $0.2 \ \Msun \le \Mzams \le 8.0 \ \Msun$ models in Figure~\ref{fig:Lnu_ratio_ZAMS_and_TAMS} have a partially degenerate He-rich core. As low-Z models have denser, hotter and more massive cores than high-Z models, the thermal plasmon neutrino contributions to \Lnu \ are larger. More massive \Mzams\ models do not develop degenerate He-rich cores, and the small dependence on the initial Z continues. The most metal-rich track decreases due to the larger mass loss.

At core He-depletion (CHeD), the 0.9\,\Msun $\le$ \Mzams $\le$ 8.0\,\Msun \ models have a partially electron-degenerate CO-rich core. The denser, hotter and more massive cores of the low-Z models means larger thermal neutrino contributions, and thus  \Lnu \ is larger in lower Z models. 

The \Mzams\,$\ge$\,60~\Msun \ models at CHeD in Figure~\ref{fig:Lnu_ratio_ZAMS_and_TAMS} show sawtooth profiles with the lowest Z models disrupting a metallicity trend. This occurs because the convective boundary mixing model, exponential overshooting \citep{herwig_2000_aa}, is based on the pressure scale height $H = P/(\rho g) \simeq k_B T / (\mu_{\rm ion} g)$, where $P$ is the pressure, $k_{\rm B}$ is the Boltzmann constant, and $g$ is the gravitational acceleration. All else being equal, a smaller Z means a smaller $\mu_{\rm ion}$, a larger $H$, and thus the chemical mixing region in low-Z models is larger than in high-Z models. If two burning shells are within $H$, they are mixed. For masses with low \Lnu, the H-shell mixes into the burning He core repeatedly. This delays core He burning until there is a homogeneous stripped CO core with a little He on the surface. By CHeD there is no H-shell to undergo CNO burning and all the $^{14}$N is depleted, ergo \Lnu \ is very low.

Overall, for fixed overshooting parameters, metal-poor models have larger amounts of chemical mixing. This is a secondary driver of changes to the thermal and nuclear reaction neutrino emission as the initial metallicity changes. Other specific examples of overshooting dominating are shown for low-mass models in Section~\ref{s.lowmass} and for high-mass models in Section~\ref{s.highmass}. The overshooting prescription may have an additional metallicity dependence that is not captured by these models.

Figure~\ref{fig:Lnu_ratio_ZAMS_and_TAMS} also compares \Enu \ at each \Mzams \ to \Enu \ of the Z\,=\,1\,\Zsun \  model on a linear scale at three evolutionary stages in the bottom three panels. At the TAMS, models across the mass spectrum reflect the Z dependence of \Lnu \ shown in the top two panels. At CHeD, the denser, hotter and more massive cores of the low-Z models, plus contributions from the conversion of $^{14}$N into $^{22}$Ne, also show larger \Enu \ with decreasing Z.

Tracks in the bottom panel of Figure~\ref{fig:Lnu_ratio_ZAMS_and_TAMS} are the same neutrino tracks in  Figure~\ref{fig:tot_energy_combined_mass} but normalized to the solar metallicity track.
The \Mzams \ range for He WD and CO WD have the metallicity signature of having had an inert, electron-degenerate core during their evolution. The ONeMg WD region shows a sawtooth pattern because these models had numerical challenges completing the propagation of their off-center, convectively bounded flame fronts to the center. The \Mzams\  region for CC events show a weak dependence of \Enu \ on Z.

\section{Low-Mass Stars}\label{s.lowmass}

Here we analyze the neutrino emission from the low-mass stellar tracks at one metallicity in Section~\ref{s.loneZ}, and then for all six metallicities in Section~\ref{s.lMetallicity}.

\subsection{One Metallicity}\label{s.loneZ}

Figure \ref{fig:hr_lowmass} shows the $0.2 \ \Msun \le \Mzams \le 7.0 \ \Msun$ tracks in a $\gamma$HRD and a $\nu$HRD for Z\,=\,1~\Zsun. The cores are progressively enriched with the ashes of H-burning as the models begin to evolve beyond the MS. The H-burning reactions increase $\mu$ and thus $\rho$ in the core. To maintain hydrostatic equilibrium the central temperature \Tc \ rises with the central density \rhoc, increasing the rate of nuclear fusion and thus \Lgamma \ and $L_{\nu}$. This slow increase of \Tc \ is reflected in the $\gamma$HRD and $\nu$HRD of Figure \ref{fig:hr_lowmass} as an increase in their respective luminosities until core H-depletion at the terminal-age main-sequence (TAMS). 

The He-rich core contracts as an H-burning shell forms and the tracks in Figure \ref{fig:hr_lowmass} evolve across both HRDs on a thermal timescale. Both \Lgamma \ and \Lnu \ increase along the RGB until core He-ignition at the tip of the RGB. All tracks that reach this point have a semi electron degenerate He core with 0.5~\Msun $\le$ \Mhe $\le$ 1.7~\Msun, and a similar $L_{\gamma}$, $L_{\nu}$, and \Teff \citep{cassisi_2013_aa, serenelli_2017_aa}. Photons from the tip of the RGB provide a standard candle distance indicator \citep{da-costa_1990_aa, lee_1993_aa, madore_2023_aa}, and offer constraints on the neutrino magnetic dipole moment \citep{capozzi_2020_aa,franz_2023_aa}.

He ignition by the triple-$\alpha$ process in the 0.9 \Msun $\leq$ \Mzams\ $\leq$ 2.1 \Msun\ tracks of Figure \ref{fig:hr_lowmass} occur off-center (on-center in the 2.1 \Msun) and under semi-electron-degenerate conditions in a helium flash \citep{thomas_1967_aa,bildsten_2012_ab,gautschy_2012_aa,serenelli_2017_aa}.  
A He burning front propagates towards the center by conduction, with burning behind the front driving convection. The helium flash and the sub-flashes that follow burn very little He; the nuclear energy generated mainly goes into lifting the electron degeneracy in the core. The last sub-flash reaches and heats the center allowing stable convective core He-burning under non-degenerate conditions. 

\begin{figure*}[!htb]
    \centering
    \includegraphics[width=7.1in]{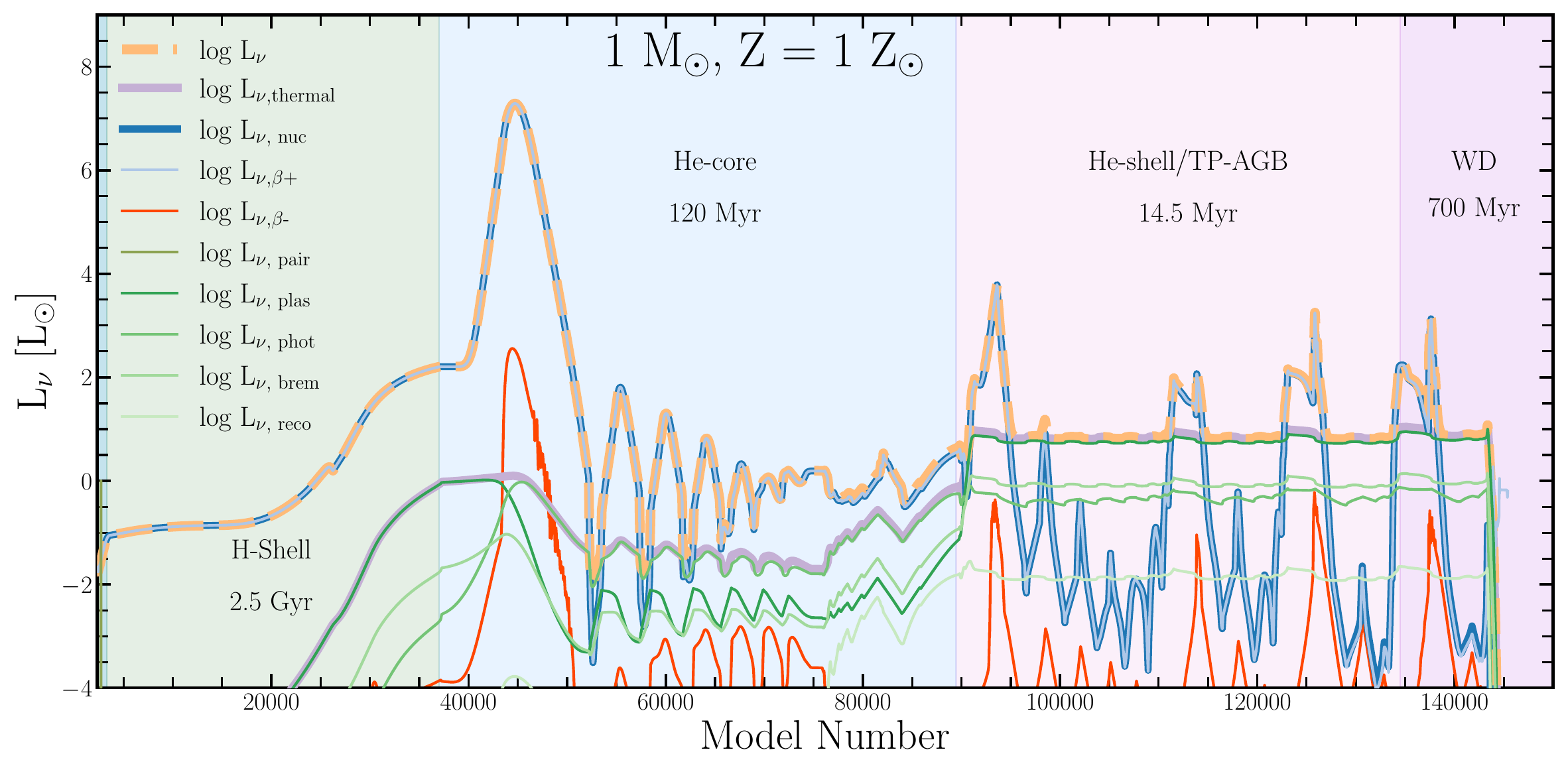}
    \includegraphics[width=7.1in]{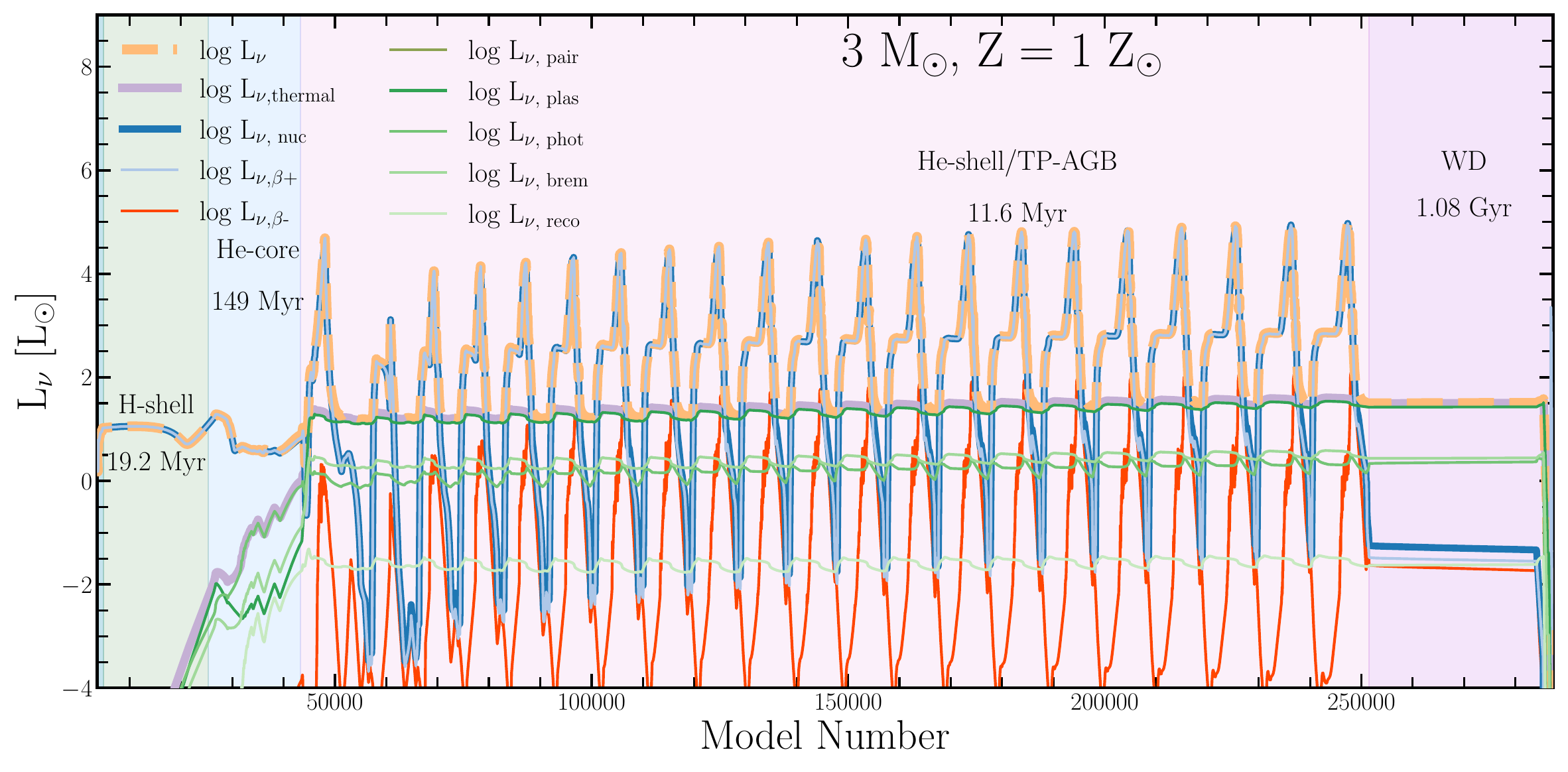}
    \caption{Components of \Lnu \ over the lifetimes of a 1~\Msun, 1~\Zsun \ model (top) and a 3~\Msun, 1~\Zsun\ model (bottom).  The x-axis is the sequential model number, a non-linear proxy for time, which begins on the left at core H-depletion and ends on the right as a cool WD at each metallicity. Phases of evolution are marked by the colored regions and the time spent in each phase is labeled. Curves show the luminosities from nuclear and thermal processes and their sub-components, and are smoothed with a 50 model moving average filter.  Luminosities are normalized to $\Lsun = 3.828 \times 10^{33}$ erg s$^{-1}$ \citep{prsa_2016_aa}. }
    \label{fig:nu_components_lowmass}
\end{figure*}

During each helium flash, a nitrogen flash also occurs from the conversion of all of the accumulated $^{14}$N to $^{22}$Ne, sharply increasing L$_{\nu}$ via $^{18}$F(,$e^{+}\nu_e$)$^{18}$O \citep{serenelli_2005_ab}. For example, Figure \ref{fig:nu_components_lowmass} shows that a \Mzams\,=\,1~\Msun, Z\,=\,1~\Zsun\ track undergoes 7 flashes. The first flash is the strongest, occurring at $M \simeq 0.2 \Msun$ and reaches $L_{\nu} \simeq 2\times10^{7}$~\Lsun\ for $\simeq 3$ days.

\begin{figure*}[!htb]
    \centering
    \includegraphics[width=7.1in]{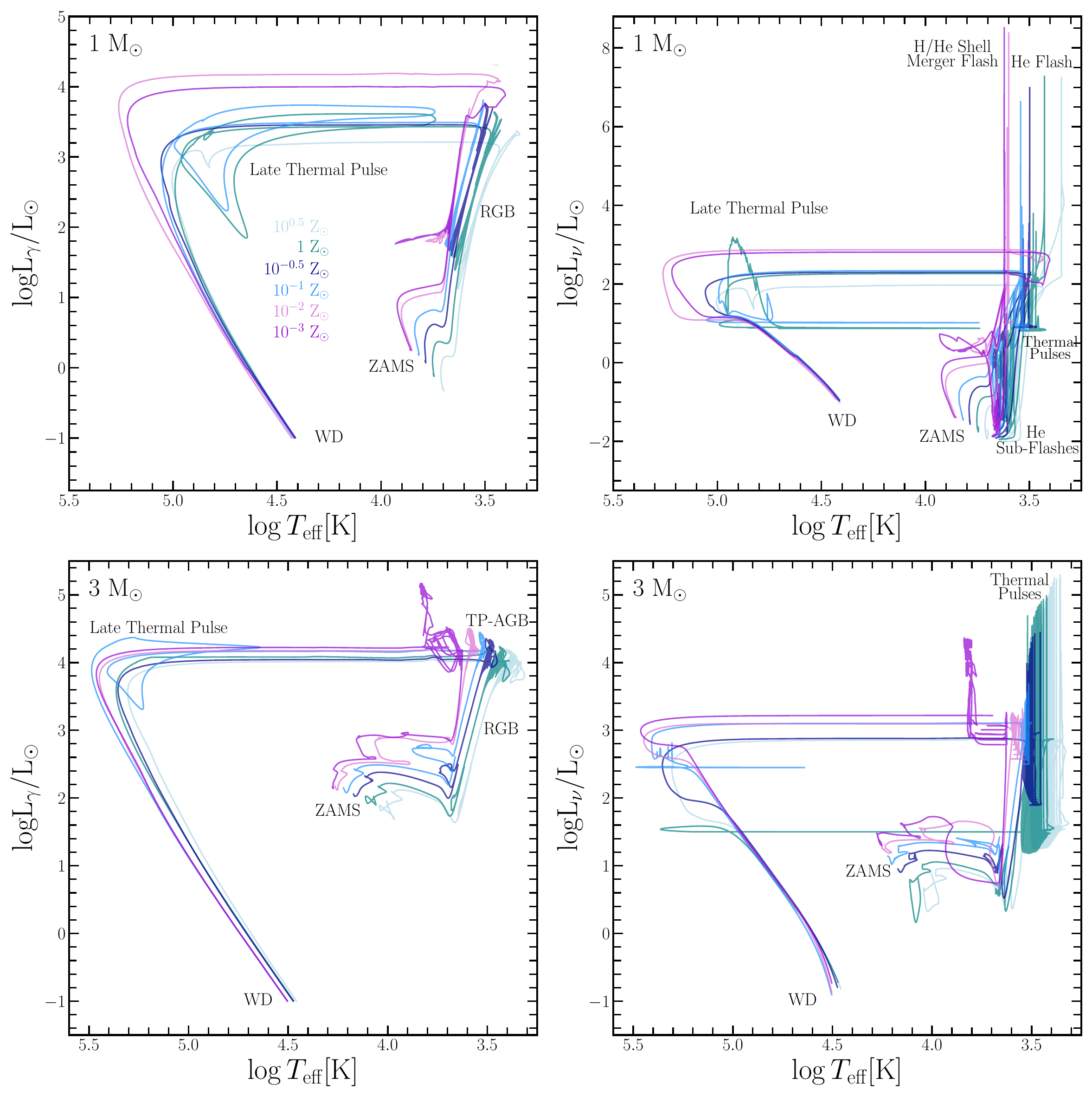}
    \caption{Tracks in a $\gamma$HRD (left panels) and $\nu$HRD (right panels) for \Mzams\,=\,1~\Msun\ (top row) and 3~\Msun\ (bottom row) across all six metallicities. Luminosities are normalized to the current solar photon luminosity \Lsun~=~3.828 $\times$10$^{33}$ erg s$^{-1}$ \citep{prsa_2016_aa}.  Metal-poor tracks are generally bluer and more luminous than metal-rich tracks.}
    \label{fig:lhrz}
\end{figure*}

Tracks with \Mzams\,$\geq$\,2.1\,\Msun\ reach a high enough \Tc \ at the tip of the RGB to ignite He in the center quiescently under non-degenerate conditions. For example, Figure \ref{fig:nu_components_lowmass} shows a \Mzams\,=\,3~\Msun,\,1~\Zsun\ track produces a smoother \Lnu  \ signature during core He burning than a \Mzams\,=\,1~\Msun,\,1~\Zsun\ track. Tracks in this mass-metallicity range also experience a blue loop \citep{hayashi_1962_aa,hofmeister_1964_aa,xu_2004_aa,zhao_2023_aa} in the $\gamma$HRD and $\nu$HRD of Figure~\ref{fig:hr_lowmass}. 

Post He ignition, the tracks in Figure \ref{fig:hr_lowmass} migrate to the horizontal branch (HB), becoming less luminous with larger  \Teff.  All the He cores have approximately the same mass, regardless of the total stellar mass, and thus about the same helium fusion luminosity $L_{\rm He}$. These stars form the red clump at \Teff\ $\simeq$ 5,000~K, $L_{\gamma} \simeq 50~\Lsun$ and $L_{\nu} \simeq 20~\Lsun$ \citep{alves_1999_aa,sarajedini_1999_aa,girardi_1999_ab,hawkins_2017_aa,wang_2021_aa}. Less massive H envelopes shift the tracks to hotter \Teff\ and smaller \Lgamma \ on the HB. This effect occurs more readily at lower Z (see \S\ref{s.lMetallicity}) with old metal-poor clusters showing pronounced HB in a $\gamma$HRD \citep{casamiquela_2021_aa,dondoglio_2021_aa}.

Core He-burning produces an electron-degenerate CO core with a semi-electron-degenerate He shell encased in a larger H-rich envelope. These AGB stars are the final stage of evolution driven by nuclear burning, characterized by H and He burning in geometrically thin shells on top of the CO core \citep{herwig_2005_aa}. Larger \Mzams \ yield super-AGB models, where an Oxygen-Neon-Magnesium (ONeMg) core forms from a convectively bounded carbon flame propagating toward the center \citep{becker_1979_aa,becker_1980_aa, timmes_1994_ab, garcia-berro_1997_aa, siess_2007_aa,denissenkov_2015_aa,farmer_2015_aa,lecoanet_2016_ab}.

During the AGB phase a thin He shell grows in mass as material from the adjacent H-burning shell is processed, causing the He shell to increase in temperature and pressure. When the mass in the He shell reaches a critical value \citep{schwarzschild_1965_aa,giannone_1967_aa,siess_2010_aa,gautschy_2013_aa,lawlor_2023_aa}, He ignition causes a thermal pulse (TP). 

For example, Figure~\ref{fig:nu_components_lowmass} shows the \Lnu \ of a 3~\Msun, 1~\Zsun\ track experiencing a series of 21 TPs, with an interpulse period of $\simeq$ 10$^5$ yr. Like the helium flash, each TP is composed of a primary flash followed by a series of weaker sub-flashes. These TP sequences appear as  spikes in the $\nu$HRD of Figure \ref{fig:hr_lowmass}. The primary flash produces the largest \Lnu \ $\simeq$ $4.6\times10^{4}$~\Lsun\ from $^{18}$F(,$e^{+}\nu_e$)$^{18}$O. The sub-flashes do not produce neutrino emissions from this process, as nearly all of the $^{14}$N is converted to $^{22}$Ne during the primary flash. The number of TPs a track undergoes is uncertain as the number is sensitive to the mass and time resolution, the stellar mass loss rate, and the treatment of convective boundaries. 

\begin{figure*}[!htb]
    \centering
    \includegraphics[width=6.0in]{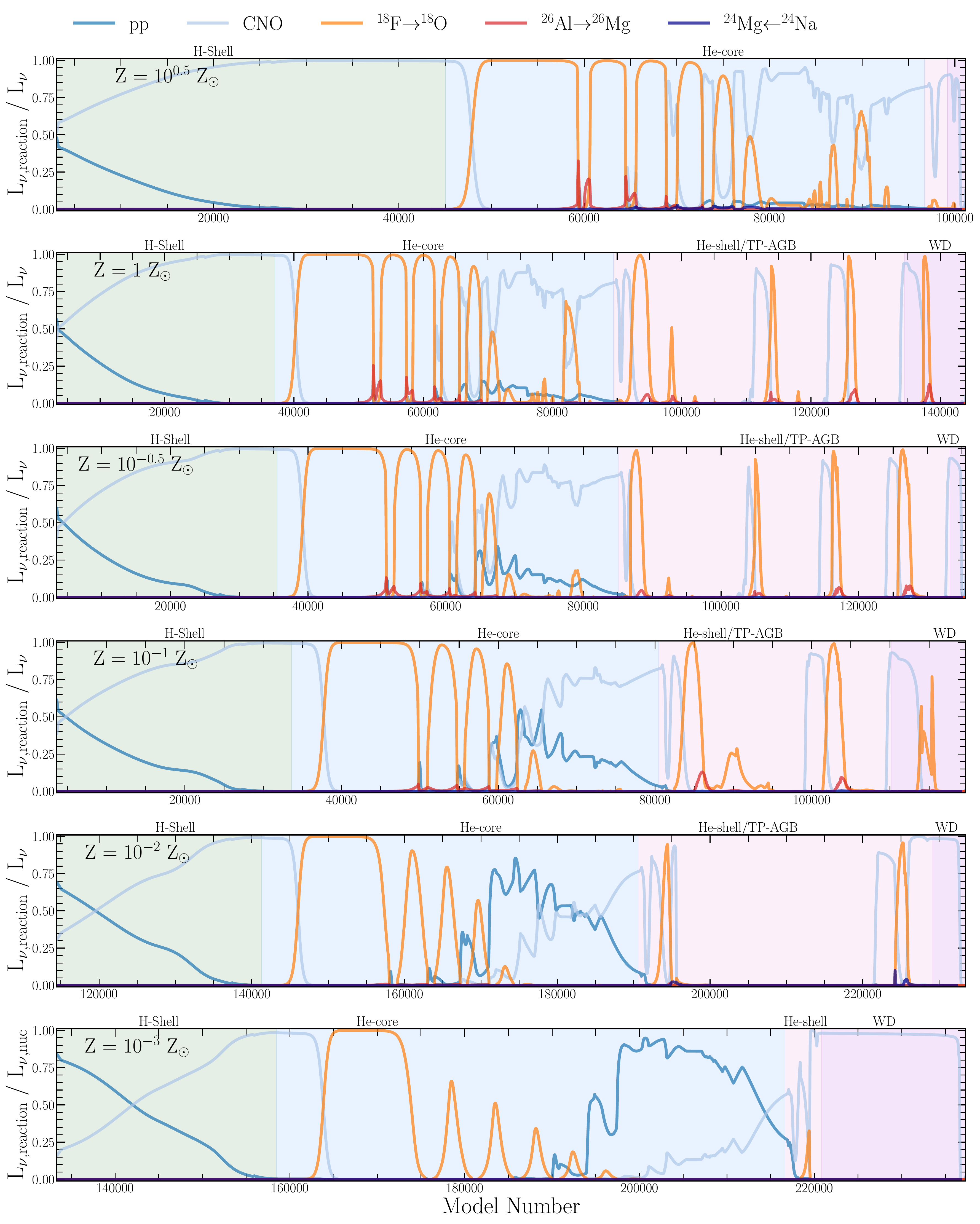}
    \caption{Components of \Lnu \ from nuclear reactions over the lifetime of a \Mzams\,=\,1 \Msun\ model for all six metallicities. The x-axis is the sequential model number, a proxy for time, beginning at core H-depletion (left) and ending as a cool WD (right). Curves are smoothed with a 50 model moving average filter. Evolutionary phases are shown by the colored regions and labelled. Reactions emitting neutrinos in the pp-chain and CNO cycles are listed in Section~\ref{s.all_oneZ}.
        }
    \label{fig:1component}
\end{figure*}

\begin{figure*}[!htb]
    \centering
    \includegraphics[width=5.6in]{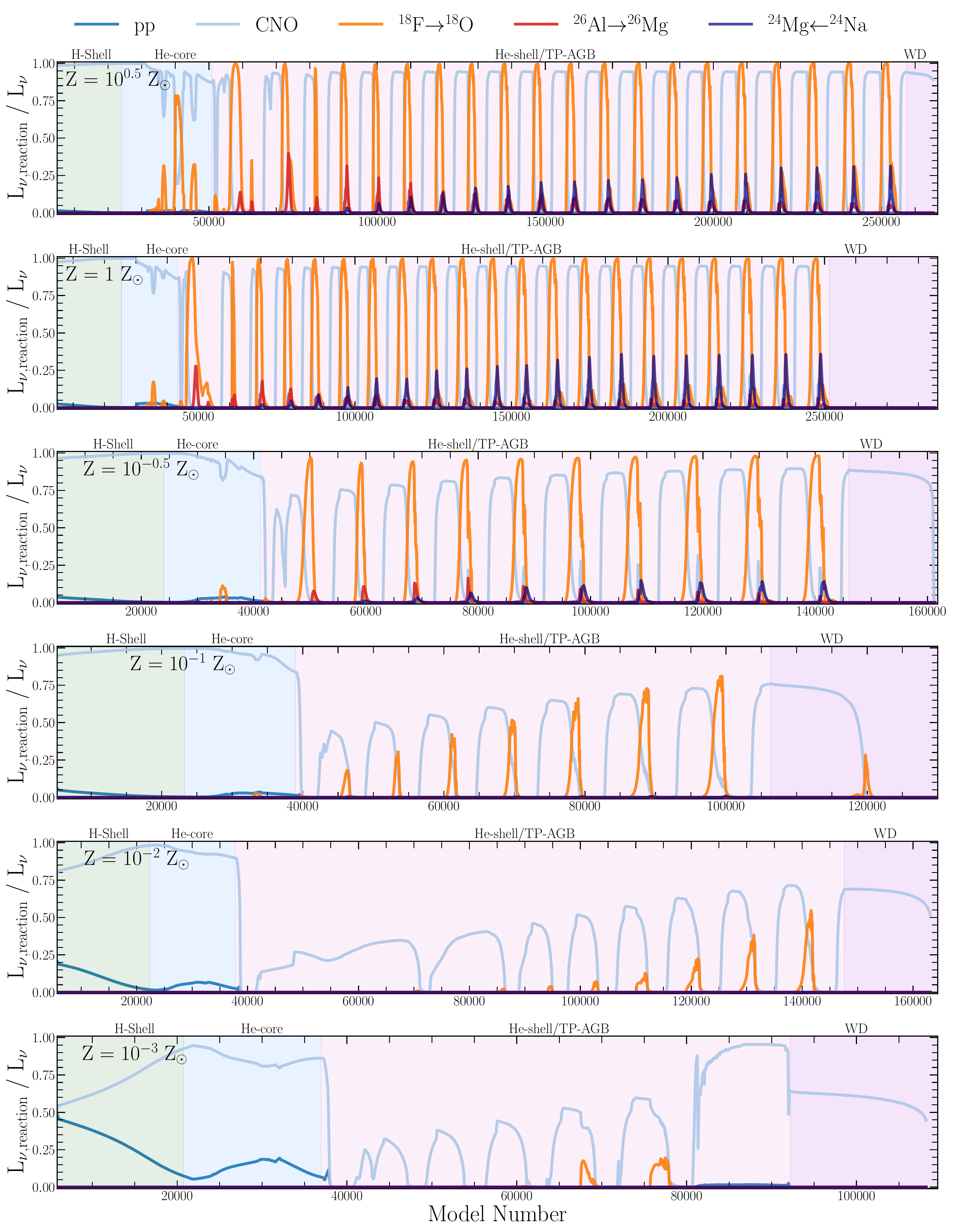}
    \caption{Same as Figure \ref{fig:1component} for  3~\Msun\ models.}
    \label{fig:3component}
\end{figure*}

\subsection{Six Metallicities}\label{s.lMetallicity}

Figure~\ref{fig:lhrz} shows the evolution of \Mzams\,=\,1~\Msun\ and 3~\Msun\ in a $\gamma$HRD and a $\nu$HRD across all six metallicities. Overall, the low-Z models show the trend of having denser, hotter and more massive cores with lower envelope opacities, larger surface luminosities and larger effective temperatures \Teff \ than the high-Z counterparts.
Features in the $\nu$HRD between core H depletion and the end of the TP-AGB phase are analyzed below.

The tracks in Figure~\ref{fig:lhrz} leave the TP-AGB phase when the envelope mass above the H and He burning shells is reduced to $\simeq$ 0.01 \Msun\ by stellar winds. All the tracks then evolve toward larger \Teff \  at nearly constant \Lnu \ and $L_{\gamma}$.  The \Mzams\,=\,1~\Msun\ and 3~\Msun\ tracks, in both the $\gamma$-HRD and $\nu$-HRD, show late TPs for some metallicities. These are the result of a strong He flash (and nitrogen flash) that occurs after the AGB phase but before the WD cooling phase \citep{iben_1983_ab,blocker_1997_aa,lawlor_2023_aa}. A candidate late TP star is V839 Ara, the central star of the Stingray Nebula \citep{reindl_2017_aa,pena_2022_aa}. The more dramatic very late TP stars, also visible in Figure~\ref{fig:lhrz}, include Sakurai's Object, V605 Aql, and perhaps HD 167362 the central star of planetary nebula SwSt 1 \citep{clayton_1997_ac,herwig_2002_aa,miller-bertolami_2007_aa,hajduk_2020_aa,lawlor_2023_aa}.

Plasmon neutrino emission then dominates the energy loss budget in Figure~\ref{fig:lhrz} for average-mass $\simeq$ 0.6 \Msun\ CO WDs with \Teff~$\gtrsim$~25,000~K \citep{vila_1966_aa,kutter_1969_aa,bischoff-kim_2018_aa}.  As the WD continues to cool, photons dominate the cooling as the electrons transition to a strongly degenerate plasma \citep{van-horn_1971_aa,corsico_2019_aa}. The tracks in Figure \ref{fig:lhrz} are arbitrarily chosen to terminate when the WD reaches $L \le 10^{-3} \Lsun$. This is sufficient \citep[see Figure 5 of][]{timmes_2018_aa} for calculating the integrated neutrino background from a simple stellar population.

Figure~\ref{fig:1component} shows the fraction of \Lnu \ from specific reaction sequences and weak reactions over the lifetime of the 1 \Msun\ models for all six metallicities. Fractions whose components do not sum to unity indicate the contribution of thermal neutrinos to \Lnu.

The green shaded regions correspond to shell H-burning.  The fraction of \Lnu \ from the CNO cycles  in this phase steadily increases with metallicity from the Z\,=\,10$^{-3}$ \Zsun \ in the bottom panel to Z\,=\,10$^{0.5}$\,\Zsun \ in the top panel.
Since the CNO nuclei catalyze H-burning, \Lgamma \ and \Lnu \ depend directly on the initial metallicity. 

The blue shaded regions represent core He-burning. In this phase, the fraction of  \Lnu \ from the $^{19}$F $\rightarrow$ $^{18}$O reaction dominates during the nitrogen flash. Neutrino emission from the H-burning pp-chain and CNO cycles appear during this phase of evolution for all six metallicities due to convective boundary mixing processes ingesting fresh H-rich material into the hotter core region.  For the Z $\ge$ 10$^{-0.5}$ \Zsun \ tracks, the convective boundary mixing processes and hotter temperatures drive the H-burning Mg-Al cycles (red curves) and the appearance of $^{26}$Al~$\rightarrow$~$^{26}$Mg between sub-flashes.

Shell He-burning and the TP-AGB phase of evolution are shown by the pink shaded regions in Figure~\ref{fig:1component}. The Z\,=\,10$^{-1, -0.5, 0}$ \Zsun \ tracks show traditional TPs, with the fractions contributing to \Lnu \ oscillating between successive TPs. Neutrino emission is initially from CNO burning before a TP, and then from $^{19}$F $\rightarrow$ $^{18}$O during the ensuing He-burning TP. 

The Z\,=\,10$^{-3}$ \Zsun \ and Z\,=\,10$^{-2}$ \Zsun \ tracks in Figure~\ref{fig:1component} do not show traditional TPs. Instead they show a single event from a merger of their H-shells and He-shells that is driven by convective boundary mixing. As analyzed in Section 3.2, this is because metal-poor models have larger chemical convective boundary mixing regions than metal-rich models for fixed overshooting parameters. The Z\,=\,10$^{0.5}$ \Zsun \ tracks in Figure~\ref{fig:1component} also do not show traditional TPs due their thinner envelopes, caused by the metallicity dependent line driven wind mass loss prescriptions removing more envelope mass ($\dot M \propto Z$). 
During the WD cooling phase (purple shaded regions) late TPS are visible in the 
Z\,=\,10$^{-1, 0}$ \Zsun \ tracks by the rise of \Lnu \ from CNO burning and subsequently $^{19}$F~$\rightarrow$~$^{18}$O.

\begin{figure*}[!htb]
    \centering
    \includegraphics[width=6.4in]{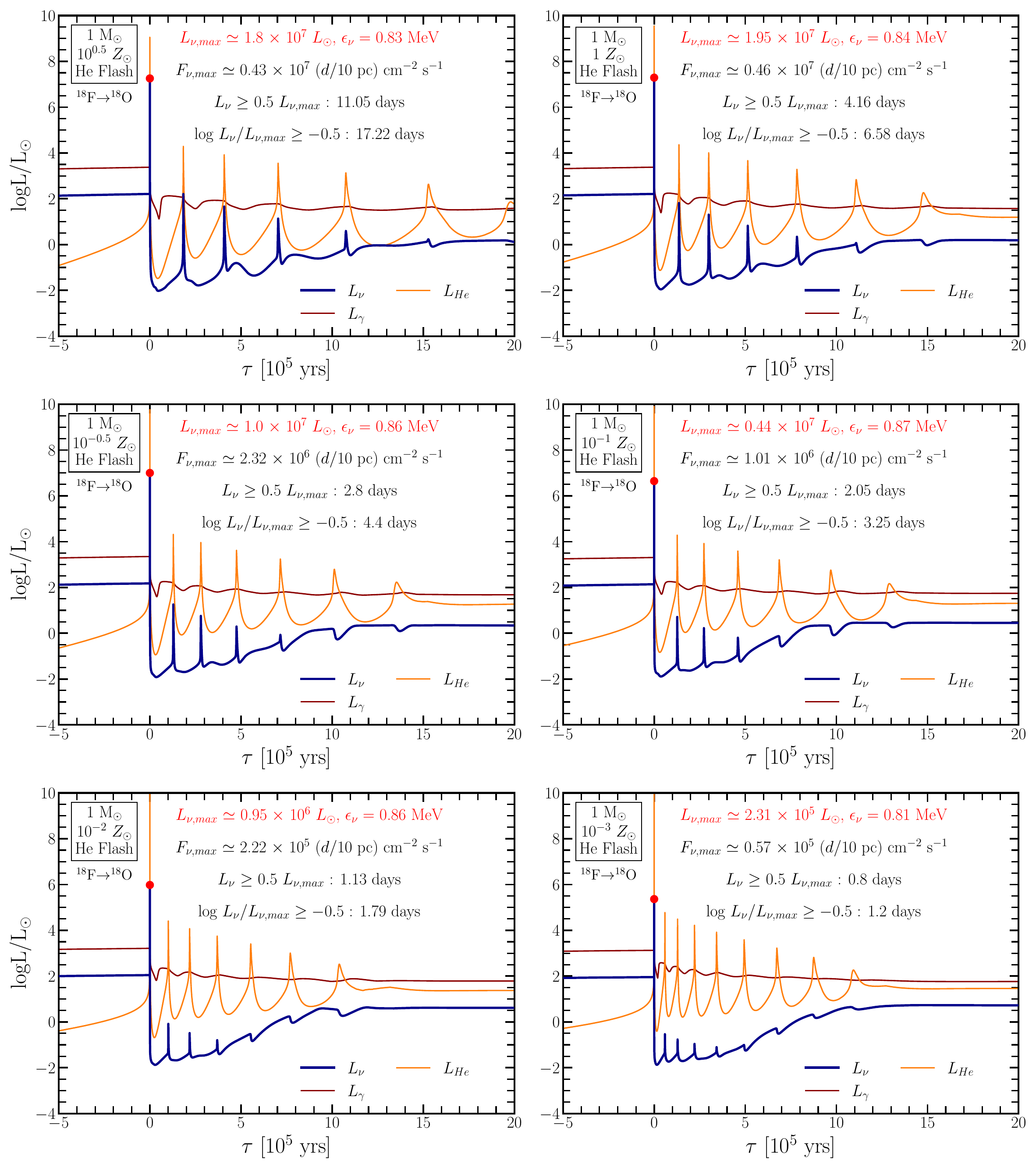}
    \caption{Neutrino targets for the nitrogen flash in 1 \Msun\ models for all six metallicities. The x-axis is the time since the first, and strongest, nitrogen flash. The y-axis is a luminosity relative to \Lsun~=~3.828 $\times$10$^{33}$ erg s$^{-1}$ \citep{prsa_2016_aa}. Colored curves show the $\nu$, $\gamma$, and He-burning luminosity. The red circle marks the maximum \Lnu \ and the red label gives the value of  $L_{\nu,{\rm max}}$ and the average neutrino energy. Labelled are the maximum flux at a distance $d$ in parsec, and the duration for the \Lnu \ to be larger than 1/2 and 1/3 $L_{\nu,\rm{max}}$. Metal-rich models have larger $L_{\nu,\rm{max}}$ and longer periods between flashes.}
    \label{fig:1flashtarget}
\end{figure*}

\begin{figure*}[!htb]
    \centering
    \includegraphics[width=6.5in]{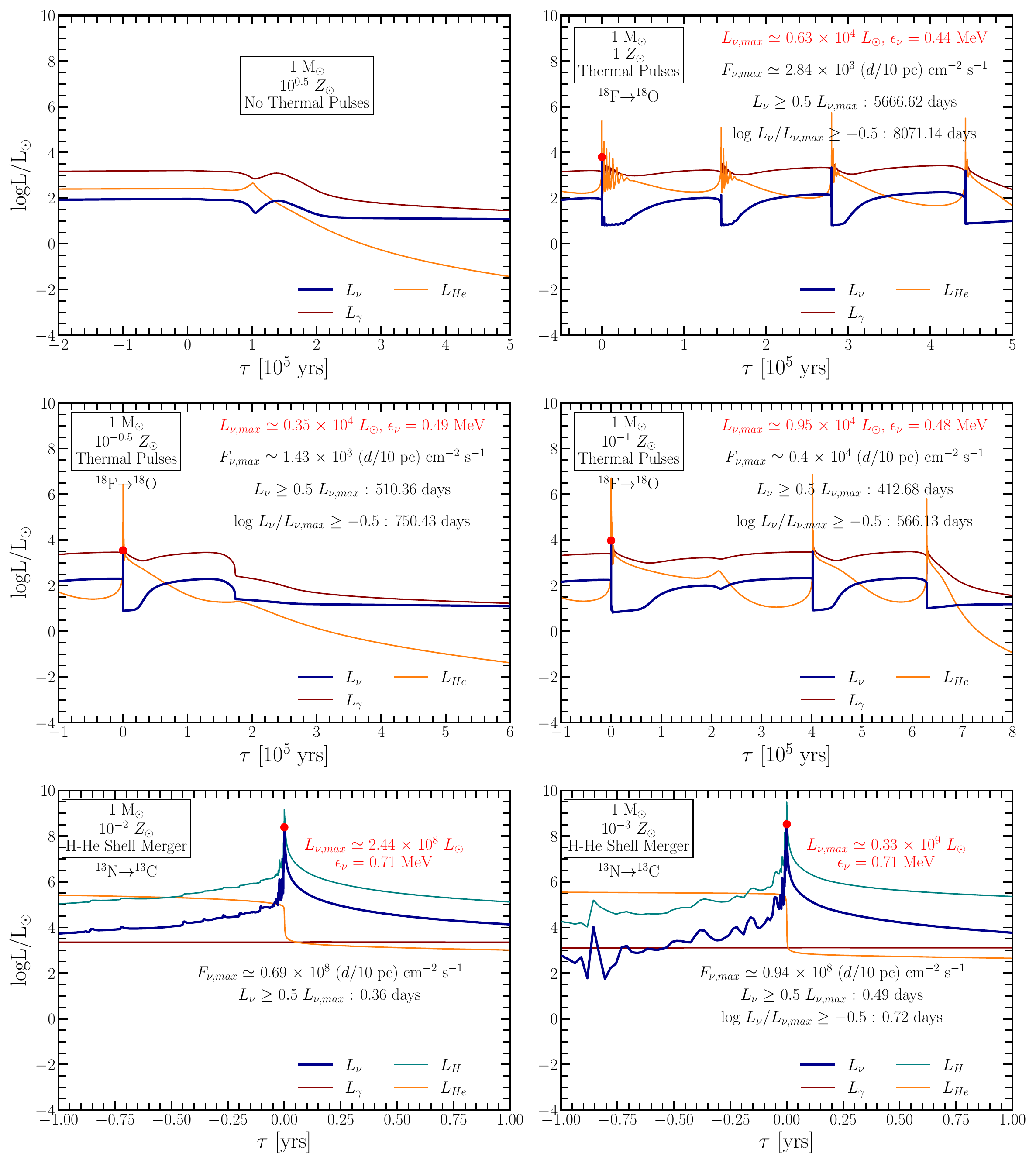}
    \caption{Same format as Figure \ref{fig:1flashtarget} but for the TP-AGB phase of evolution. }
    \label{fig:1tptarget}
\end{figure*}

\begin{figure*}[!htb]
    \centering
    \includegraphics[width=6.4in]{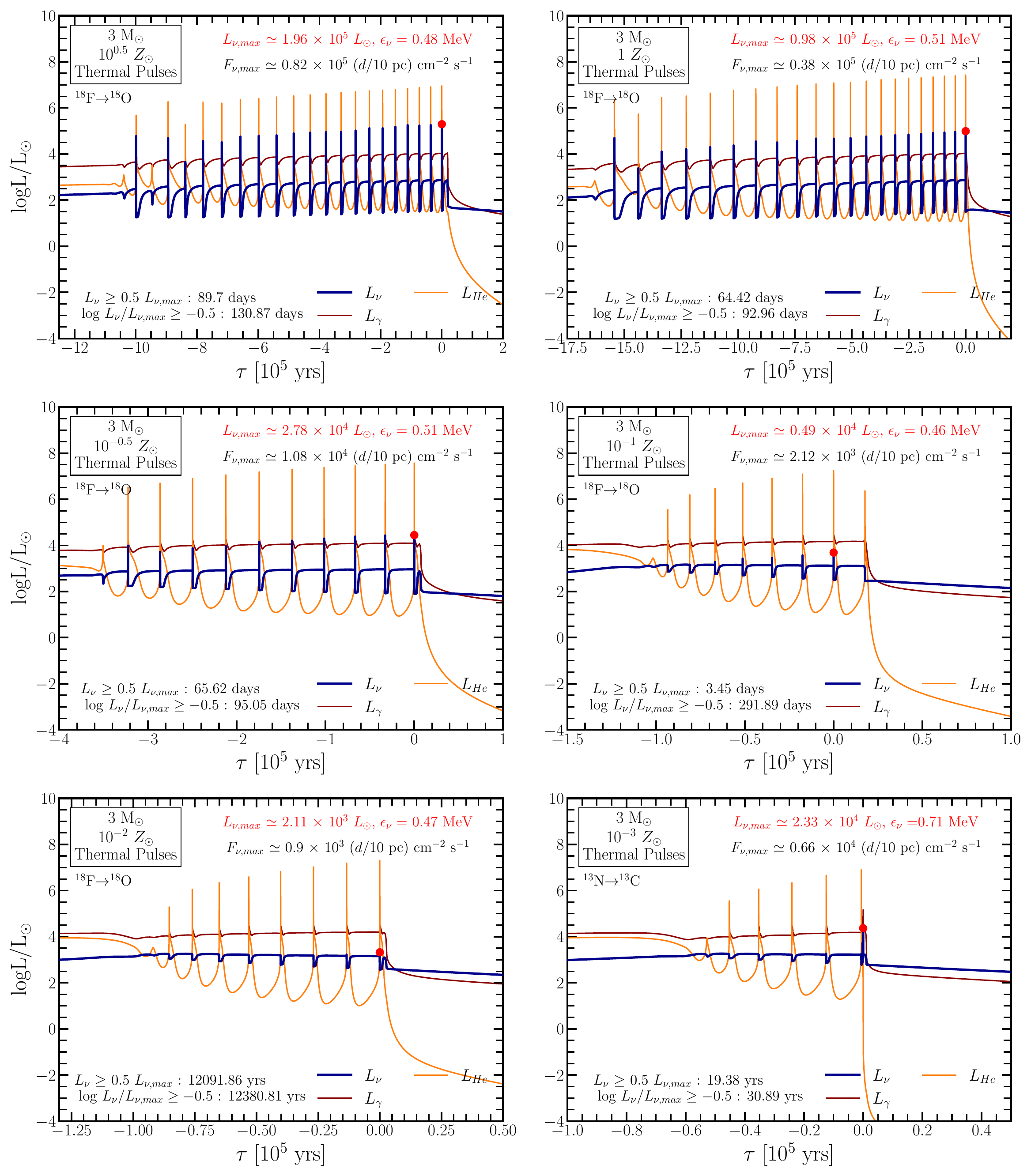}
    \caption{Same format as Figure~\ref{fig:1tptarget} but for 3~\Msun\ models across all six metallicities. }
    \label{fig:3tptarget}
\end{figure*}

Figure~\ref{fig:3component} is the same as Figure~\ref{fig:1component} but for the lifetime of the 3 \Msun\ models for all six metallicities.  The fraction of \Lnu \ from CNO processing during shell H-burning (green regions) is larger for the 3~\Msun\ tracks than the corresponding 1~\Msun\ tracks of Figure~\ref{fig:1component} at all metallicities. Core He-burning (blue shaded regions) proceeds smoothly under non-degenerate conditions at all metallicities. The spikes from $^{19}$F~$\rightarrow$~$^{18}$O in the Z\,=\,10$^{0.5}$\,\Zsun \ track during core He-burning are due to overshooting injecting fresh H-rich fuel into the core. Shell He-burning and the TP-AGB phase of evolution (pink regions) show a trend of stronger and more numerous TPs as the metallicity increases from Z\,=\,10$^{-3}$ \Zsun \ to Z\,=\,10$^{0.5}$ \Zsun. Hotter temperatures in the 3~\Msun\ models cause neutrino emission from $^{26}$Al~$\rightarrow$~$^{26}$Mg during the H burning Ne-Na cycle (red curves) and from the inverse beta decay $^{24}$Na~$\rightarrow$~$^{24}$Mg reaction (purple curves). While $^{24}$Na is not part of the H burning Mg-Al cycle, this isotope is synthesized at low abundance levels during the Mg-Al cycle. A late TP occurs during the WD cooling phase (purple regions) for the Z\,=\,10$^{0, -1}$ \Zsun \ tracks.

Figure \ref{fig:1flashtarget} shows $L_{\nu}$, $L_{\gamma}$, and the He-burning luminosity \LHe \ during the nitrogen flash in 1 \Msun\ models. 
Across all metallicities the first flash has the largest \Lnu \ and \LHe \ with \LHe \ $>$ $L_{\nu}$. The  maximum neutrino luminosity $L_{\nu,{\rm max}}$, marked by the red circles and labels, spans $\simeq$\,2 orders of magnitude as the initial metallicity varies from Z\,=\,10$^{-3}$ \Zsun \ to Z\,=\,10$^{0.5}$ \Zsun. Note $L_{\nu,{\rm max}}$ is larger for the Z\,=\,1 \Zsun\  model than the Z\,=\,10$^{0.5}$ model. This is due to mass loss. If the metallicity was 10$^{0.3}$ \Zsun, then $L_{\nu,{\rm max}}$ at the He flash would be larger than the Z\,=\,1 \Zsun \ model.
At  Z\,=\,10$^{0.5}$ \Zsun, mass-loss hampers the strength of the He flash. 
The Z\,=\,10$^{0.5}$ \Zsun \ model has $M$\,=\,0.6 \Msun \ at the onset of He-flash, while the Z\,=\,1~\Zsun \ model has $M$\,=\,0.66 \Msun.  The smaller shell burning temperatures is sufficient to weaken $L_{\nu,{\rm max}}$.
Note the duration of the peak in the Z\,=\,10$^{0.5}$~\Zsun \ model is significantly longer than in the Z\,=\,1 \Zsun \ model, ensuring more neutrinos are produced overall from the larger $^{14}$N reservoir, but with a  $L_{\nu,{\rm max}}$ of similar magnitude.

Figure \ref{fig:1flashtarget} shows the average neutrino energy at $L_{\nu,{\rm max}}$ is insensitive to the initial Z. The neutrino fluxes at $L_{\nu,\rm{max}}$ span $\simeq$\,2 orders of magnitude across metallicity and can serve as target values for neutrino observations of the nitrogen flash. The duration where \Lnu~$\ge$~1/2 $L_{\nu,\rm{max}}$ increases steadily from $\simeq$\,0.8 days at Z\,=\,10$^{-3}$~\Zsun \ to $\simeq$ 11 days at Z\,=\,10$^{0.5}$~\Zsun . The duration where \Lnu~$\ge$~1/3 $L_{\nu,\rm{max}}$ increases from $\simeq$\,1.2 days at Z\,=\,10$^{-3}$~\Zsun \ to $\simeq$\,17 days at  Z\,=\,10$^{0.5}$~\Zsun. In addition, the time period between sub-flashes increases from $\simeq$ 10$^{5}$~yr at Z\,=\,10$^{-3}$~\Zsun \ to $\simeq$  2$\times$10${^5}$~yr at Z\,=\,10$^{0.5}$~\Zsun \ while the number of sub-flashes ranges between 8 at the lowest initial Z to 5 at the largest initial Z.

\begin{figure*}[!htb]
    \centering
    \includegraphics[width=6.2in]{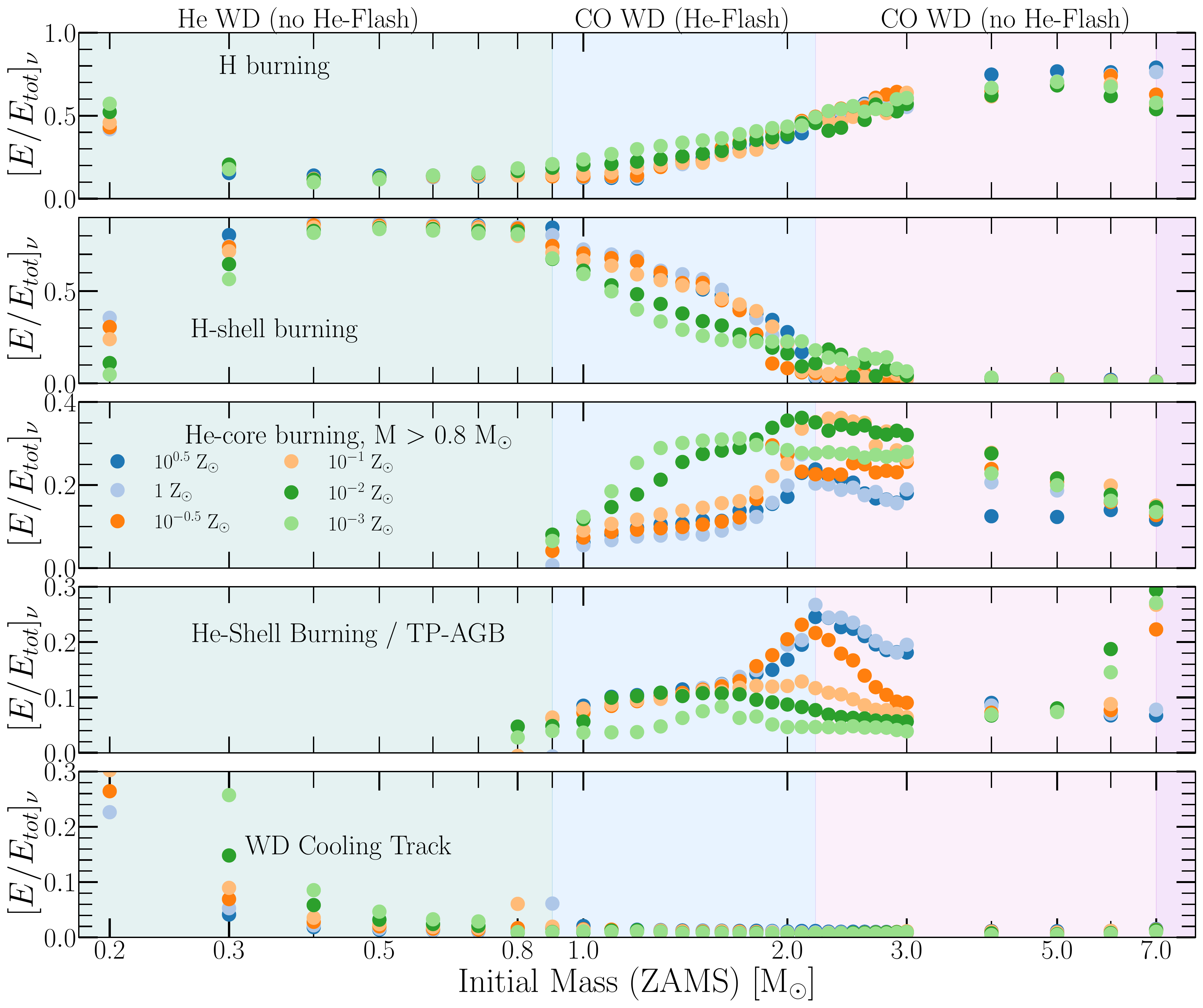}
    \caption{Fraction of \Enu\ emitted at different phases of evolution for all six metallicities (colored circles). From top to bottom, the panels show $[E/E_{\rm tot}]_{\nu}$ for core H-burning, shell H-burning, core He-burning prior to any TP-AGB phase, He-burning through the TP-AGB phase, and during the WD cooling phase.}
    \label{fig:low_e_vs_m}
\end{figure*}

\begin{figure*}[!htb]
    \centering
    \includegraphics[width=7.1in]{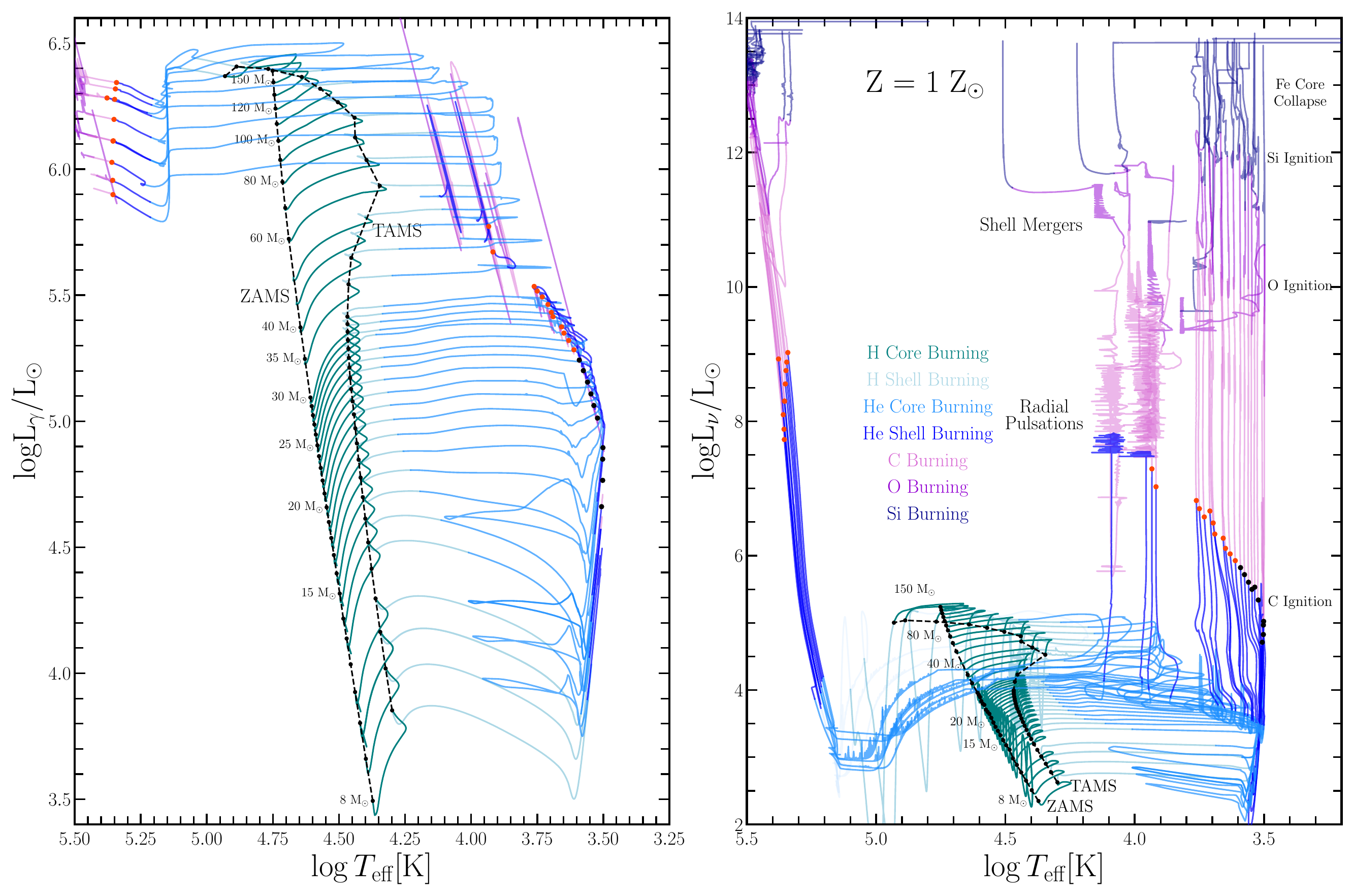}
    \caption{Tracks in a $\gamma$HRD (left panels) and a $\nu$HRD (right panels) for Z\,=\,1\,\Zsun \ over \Mzams\,=\,8--150 \Msun. Tracks are colored by evolutionary phase and labeled. Black circles mark convective core C-burning, red circles mark radiative core C-burning. Luminosities are normalized to $\Lsun = 3.828 \times 10^{33}$ erg s$^{-1}$ \citep{prsa_2016_aa}.}
    \label{fig:high_mass_hr}
\end{figure*}

Figure \ref{fig:1tptarget} shows $L_{\nu}$, $L_{\gamma}$, and \LHe \ during the TP-AGB phase of evolution in 1 \Msun\ models for all six metallicities. As discussed for Figure~\ref{fig:1component}, the tracks for the lowest initial Z show a single H-shell and He-shell merger event instead of a traditional TP. For these models \Lnu \ is dominated by $^{13}$N~$\rightarrow$~$^{13}$C from non-equilibrium hot CNO cycle burning.  At the peak of the merger $T$\,$\simeq$\,2$\times$10$^{8}$~K and $\rho$\,$\simeq$\,10$^4$ g cm$^{-3}$. At these conditions the first half of the CNO cycle, $^{12}$C($p$,$\gamma$)$^{13}$N(,$e^{+}\nu$)$^{13}$C($p$,$\gamma$)$^{14}$N, is sufficiently energetic to cause a rapid expansion that self-quenches the second half of the CNO cycle, $^{14}$N($p$,$\gamma$)$^{15}$O(,$e^{+}\nu$)$^{15}$N($p$,$\alpha$)$^{12}$C. For example, the stellar radius $R$ of the Z\,=\,10$^{-3}$~\Zsun \ model rapidly increases from 68 \Rsun \ to 465 \Rsun \ during the merger and the number of reactions per second from $^{13}$N~$\rightarrow$~$^{13}$C is $\simeq$\,3 orders of magnitude larger than from $^{15}$O~$\rightarrow$~$^{15}$N. Thus, these 1~\Msun \ low-Z models do not undergo a TP  because a violent shell merger causes the model to quickly lose most of the H envelope. These mergers, driven by convective boundary mixing, produce the largest $L_{\nu,{\rm max}}$ events over the entire evolution. They are also prominent and labeled in the $\nu$HRD of Figure~\ref{fig:lhrz}. The Z\,=\,10$^{0.5}$~\Zsun \ track also does not show TPs due to their thinner envelopes from wind mass loss. For the other metallicities,  $L_{\nu,{\rm max}}$ during the TPs is $\simeq$\,3 orders of magnitude smaller than $L_{\nu,{\rm max}}$ from the nitrogen flash shown in Figure~\ref{fig:1flashtarget}.

Figure \ref{fig:3tptarget} shows $L_{\nu}$, $L_{\gamma}$, and \LHe \ during the TP-AGB phase in the 3 \Msun\ models. The number of \Lnu \ peaks (6 to 21), the \Lnu \ peaks (2$\times$10$^{3}$~\Lsun\ to 2$\times$10$^{5}$~\Lsun), and time between \Lnu \ peaks (2$\times$10$^{4}$ yr to 4$\times$10$^{4}$ yr) increase with Z, with evidence of saturation by Z\,=\,1~\Zsun. Each successive TP releases more nuclear energy, thus $L_{\nu,{\rm max}}$ occurs at the end of the tracks (red circles and labels) across all metallicities. The Z\,=\,10$^{-3}$~\Zsun \ model has a larger $L_{\nu,{\rm max}}$ than the Z\,=\,10$^{-2}$~\Zsun \ model due to $^{13}$N~$\rightarrow$~$^{13}$O (instead of the usual $^{18}$F $\rightarrow$ $^{18}$O) from a shell merger that is driven by convective boundary mixing.

Figure \ref{fig:low_e_vs_m} compares the fraction of the total energy emitted by neutrinos at five phases of evolution across the mass-metallicity plane. Models with 0.2~\Msun\,$\le$\,\Mzams\,$\le$\,0.8~\Msun \ emit $\simeq$\,80\% of their neutrinos during shell H-burning (second panel) with a slight trend towards high-Z tracks making larger contributions than low-Z tracks. A $\simeq$\,10\%  contribution originates from  core H-burning (top panel), and a $\simeq$\,10\% contribution occurs during the He WD cooling phase (bottom panel). These models do not go through shell He-burning phase, as indicated by the empty region in the fourth panel, and the shorter tracks in the $\gamma$HRD and $\nu$HRD of Figure~\ref{fig:lhrz}. 

Models whose final fate is a CO WD emit $\simeq$\,20--80\% of their neutrinos during core H-burning, $\simeq$\,20--40\% during core He-burning, and $\simeq$\,10--30\% during the TP-AGB phase. The percentages increase with \Mzams, and with Z for more massive models.

\begin{figure*}[!htb]
    \centering 
    \includegraphics[width=7.1in]{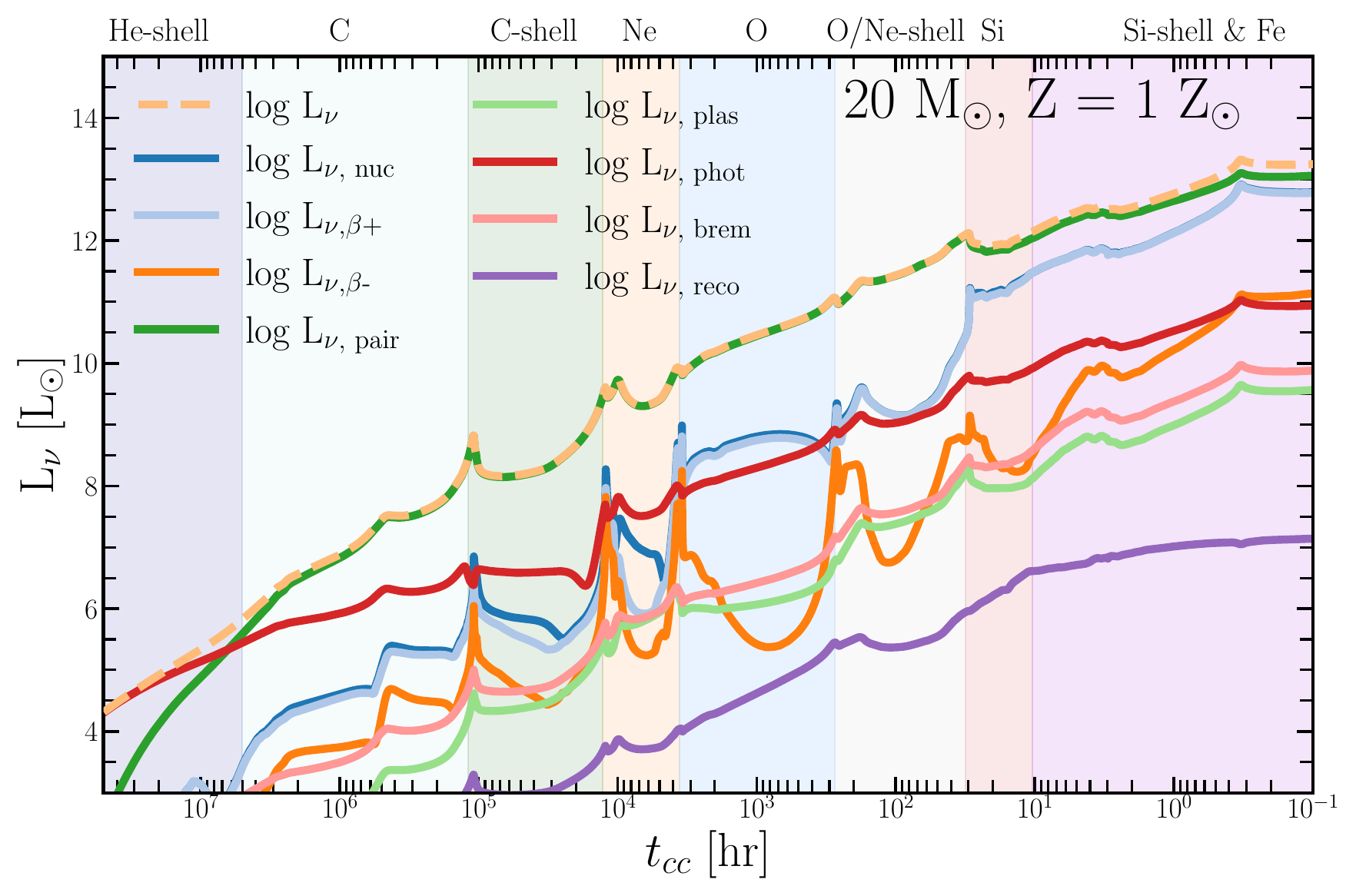}
    \caption{Components of \Lnu \ for each phase of evolution of a \Mzams\,=\,20 \Msun, 1 \Zsun model.  The x-axis is the time to the onset of CC. Curves show the contributions from nuclear and thermal neutrinos, and are smoothed with a 50 timestep moving average filter. Phases of evolution are shown by the colored regions and labeled above the plot. Boundaries between phases are defined by the ignition of the next fuel source, when the central mass fraction of the next fuel source decreases by $10^{-3}$.}
    \label{fig:fig5}
\end{figure*}

\begin{figure*}[!htb]
    \centering 
    \includegraphics[width=7.1in]{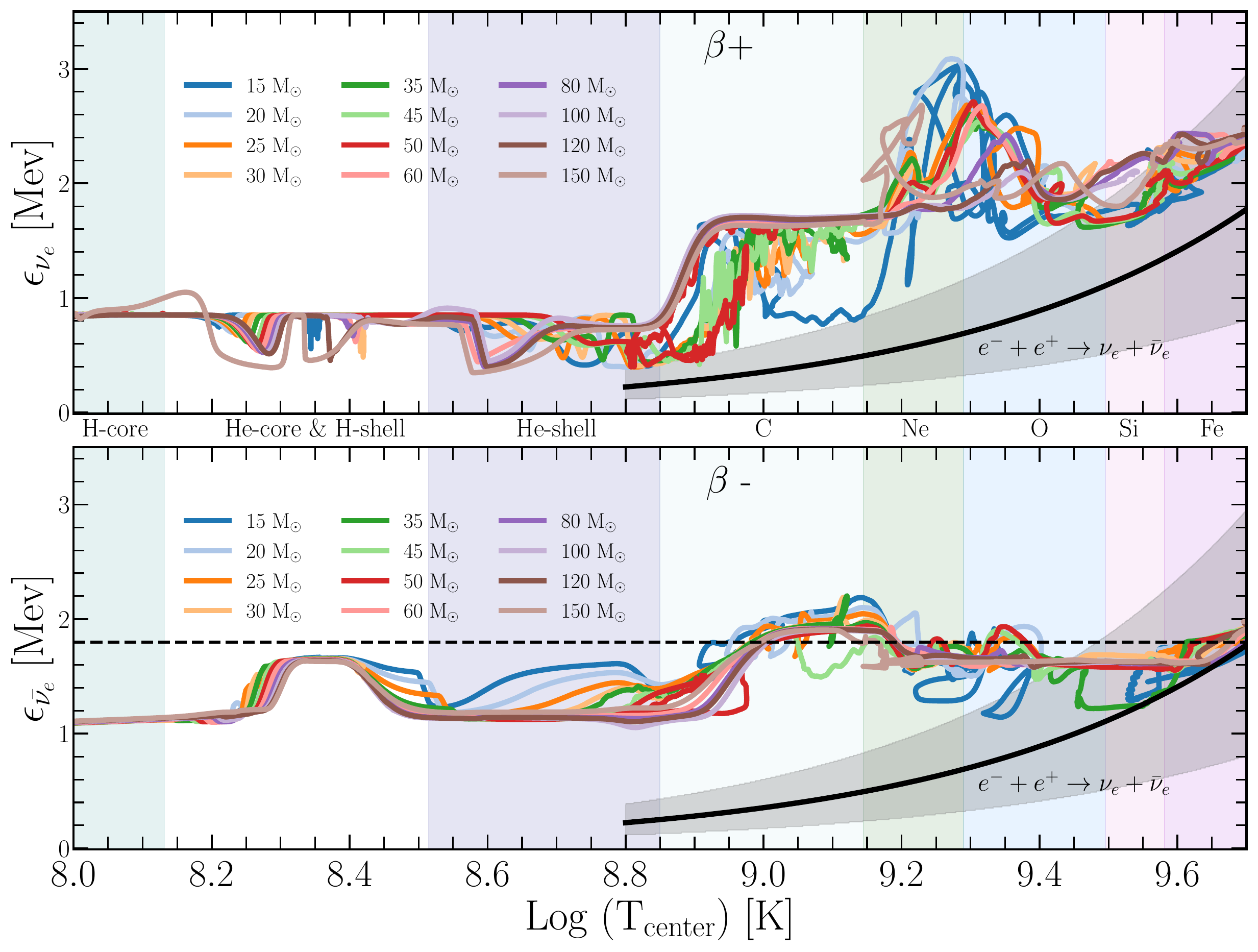}
    \caption{Average electron neutrino energy for beta decays (top), and average electron anti-neutrino energy for inverse-beta decays (bottom) for the 1 \Zsun \ models across a range of \Mzams. Curves are smoothed with a 50 timestep moving average filter. The average pair-neutrino energy is shown by the black curve, with the gray band giving the lower 10\% and upper 90\% of pair-neutrino energies. Phases of evolution are shown by the colored panels and labeled. The horizontal dashed line shows a representative $\simeq$\,1.8~MeV detection threshold to inverse beta decay of current neutrino detectors \citep[e.g.,][]{simpson_2019_aa, harada_2023_aa}. The average electron neutrino and anti-neutrino energies are approximately independent of \Mzams.}
    \label{fig:fig6}
\end{figure*}

\begin{figure*}[!htb]
    \centering
    \includegraphics[width=7.1in]{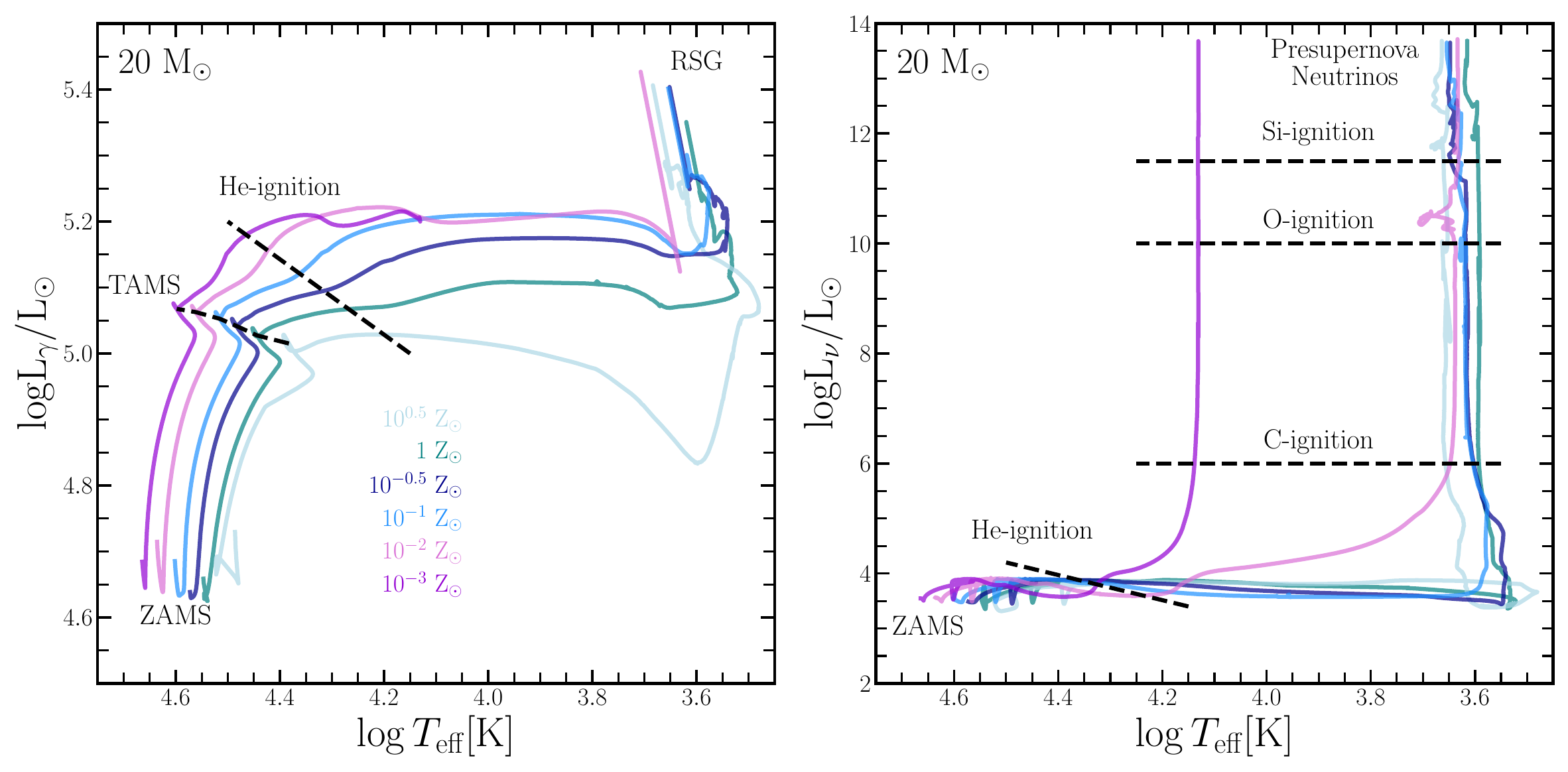}
    \caption{ Tracks in a $\gamma$HRD (left) and $\nu$HRD (right) from ZAMS to the onset of CC for the \Mzams\,=\,20~\Msun\ models across all six metallicities. Approximate locations of evolutionary phases are labeled or marked with a black dashed line. Luminosities are normalized to \Lsun~=~3.828 $\times$10$^{33}$ erg s$^{-1}$ \citep{prsa_2016_aa}.}
    \label{fig:high_mass_fighr}
\end{figure*}

\section{High-mass Stars}\label{s.highmass}

We present features of the neutrino emission from high-mass models for one metallicity in Section~\ref{s.honeZ}, and then for all six metallicities in Section~\ref{s.hMetallicity}.

\subsection{One Metallicity}\label{s.honeZ}

Tracks from the ZAMS to the onset of CC for the 8\,\Msun\,$\le$\,\Mzams\,$\le$\,150 \Msun\ models in a $\gamma$HRD and $\nu$HRD is shown in Figure \ref{fig:high_mass_hr}.  All tracks evolve at roughly constant \Lgamma \ and  \Lnu \ during core H-burning and He-burning as the tracks evolve from ZAMS to cooler \Teff. Neutrinos from the CNO cycles and $^{14}$N\,$\rightarrow$\,$^{22}$Ne power \Lnu \ through these phases of evolution. From CHeD onwards, the dominance of \Lnu \ from the core over \Lgamma \ from the surface results in a rapid reduction in evolutionary timescales from years to hours to seconds \citep{fowler_1964_aa,deinzer_1965_aa,woosley_2002_aa}. This escalating rapidity  appears in the $\nu$HRD as the nearly vertical curves at approximately constant \Teff. 

For \Mzams\,$\lesssim$\,50\,\Msun, the nearly vertical tracks at cooler \Teff \ in the $\nu$HRD end their lives as red supergiants (RSG). The \Mzams\,$\gtrsim$\,50\,\Msun\, models evolve through the advanced stages at increasingly hotter \Teff\, with thinner H envelopes, until wind driven mass-loss strips the H-envelope, creating a Wolf–Rayet model. The nearly vertical tracks at hotter \Teff \ in the $\nu$HRD end their lives as a blue supergiants.  
This transition mass is the Humphrey-Davidson limit in our models \citep{humphreys_1979_aa,davies_2018_aa,davies_2020_aa,Sabhahit_2021_aa}.  The conversion of a mass limit to a luminosity limit depends on assumptions. For example, \citet{Sabhahit_2021_aa} adopt the luminosity limit as the luminosity above which a massive star model spends $<$5\% of it's lifetime or above the luminosity limit while the model is a yellow/red supergiant.
This transition mass is sensitive to the mass and time resolution,  mass-loss prescription, and treatment of super adiabatic convection in the outer envelope \citep{Sabhahit_2021_aa}.

Another feature in the $\nu$HRD of Figure \ref{fig:high_mass_hr} is the radial pulsations in the  35\,\Msun\,$\lesssim$\,\Mzams\,$\lesssim$\,50\,\Msun\  tracks that develop during He shell or C-burning, models with thin H envelopes, and 3.9\,$\lesssim$\,log(\Teff)\,$\lesssim$\,4.1. 

C-burning sets the entropy for the continued evolution to CC, by proceeding either convectively or radiatively \citep{murai_1968_aa,arnett_1972_aa,lamb_1976_aa}. If the energy released by nuclear reactions is slightly larger than pair production neutrino losses, then net energy produced is transported by convection \citep[e.g.,][]{cristini_2017_aa}.  Otherwise, the core burns carbon radiatively in balanced power \citep{woosley_2002_aa,el-eid_2004_aa,limongi_2018_aa}, where the mass averaged nuclear energy release nearly balances the mass averaged neutrino losses.
For Z\,=\,1 \Zsun, tracks for \Mzams $\leq$ 20 \Msun\ burn carbon convectively (black circles in Figure \ref{fig:high_mass_hr}) and tracks with \Mzams\,$\geq$\,21~\Msun\ burn carbon radiatively (red circles in Figure \ref{fig:high_mass_hr}). 

The decrease in entropy from thermal neutrino emission that occurs during convective core C-burning is missing during radiative core  C-burning \citep{weaver_1993_aa}. For the \Mzams $\geq$ 21 \Msun\ tracks that undergo radiative C-burning, the subsequent burning phases occur at higher entropy, $s \propto T^3/\rho \propto (M/\Msun)^2$, at higher temperatures and lower densities. The larger entropy, in turn, drives shallower and more extended density gradients, larger effective Chandrasekhar masses at core-collapse, smaller compactness parameters, and thus are more challenging to explode as CC events \citep{woosley_1986_aa,nomoto_1988_aa,Sukhbold_2014_aa,sukhbold_2016_aa,sukhbold_2018_aa,limongi_2018_aa,Sukhbold_2020_aa,Burrows_2021_n}.  This entropy bifurcation at C-burning may seed a bimodal compact object distribution for single stars that undergo convective C-burning forming one peak in the compact object initial mass function (neutron stars) and single stars that undergo radiative  C-burning forming a second peak (black holes) \citep[e.g.,][]{timmes_1996_ac,heger_2003_aa,zhang_2008_ab,piro_2017_aa,sukhbold_2018_aa,vartanyan_2018_aa,takahashi_2023_aa}.

In the terminal phases $Y_{e}$ and $\mu$ act as guides to the evolution and culminating fate. A dwindling $Y_{e}$, catalyzed by electron captures, hastens the core's contraction and amplifies energy depletion through neutrino emissions, thereby altering the core's structural equilibrium. Concurrently, as $Y_{e} \propto 1/\mu$, an ascending $\mu$ signifies a shift towards fusing isotopically heavier nuclei, requiring ever larger core temperatures and densities to maintain hydrostatic equilibrium. 

In addition, dynamical large-scale mixing on nuclear burning timescales can occur, as can mergers between the He, C, Ne, O, and Si shells. These shell mergers are sensitive to the mixing scheme adopted and particularly the treatment of convective boundary mixing across shell boundaries \citep[e.g.,][]{ritter_2018_aa,fields_2021_aa}. An approximate location of these shell mergers is labeled in the $\nu$HRD of Figure \ref{fig:high_mass_hr}. Strong coupling between nuclear burning and turbulent convection develop during late O-burning which requires 3D simulations to establish the fidelity of the 1D convection approximations \citep{meakin_2007_ab,couch_2015_aa,muller_2017_aa,fields_2020_aa, fields_2021_aa}. As the Fe core approaches its effective Chandrasekhar mass, electron capture and photodisintegration of nuclei drive the onset of CC. 

Figure \ref{fig:fig5} shows the components L$_{\nu}$ for each phase of evolution in the \Mzams\,=\,20\,\Msun\ Z\,=\,1\,\Zsun \ model,
from shell He-burning on the left to CC on the right. After CHeD the CO core cools and contracts as a convective He-burning shell forms. The first panel on the left shows the energy budget becomes increasingly dominated by photoneutrino production with \Lnu\,$\simeq$\,10$^{5}$ \Lsun.

At $t_{\rm cc}$\,$\simeq$\,574 yr, carbon ignites with \Lnu\,$\simeq$\,10$^{6}$ \Lsun\, and the energy budget becomes dominated by pair annihilation (second panel) . Thermal neutrinos from plasmon decay, bremsstrahlung, and recombination have luminosities several orders of magnitude smaller. 

At $t_{\rm cc}$\,$\simeq$\,13.6 yr the C-shell ignites (third panel), with a sharp increase in 
$L_{\nu,{\rm nuc}}$\,=\,$L_{\nu,\beta+}$\,+\,$L_{\nu,\beta^-}$\,$\simeq$\,10$^{6.7}$~\Lsun. 
At $t_{\rm cc}$\,$\simeq$\,1.5 yr Ne ignites (fourth panel) also with a second sharp increase in 
$L_{\nu,\beta-}$ and $L_{\nu,{\rm nuc}}$/\Lnu\,$\simeq$\,5\%. 
At $t_{\rm cc}$\,$\simeq$\,0.5 yr (fifth panel) core O ignites. Convection mixes some of the Ne-shell into the core inducing a third spike in $L_{\nu,\beta-}$ and $L_{\nu,{\rm nuc}}$/\Lnu\,$\simeq$\,15\%. 
At $t_{\rm cc}$\,$\simeq$\,11.5 day (sixth panel) the O-Neon shell ignites, producing a fourth spike with L$_{\nu,nuc} \sim 10^{9}$ and L$_{\nu} \sim 10^{11}$, followed shortly by a subdued fifth spike marking the depletion of the Ne-shell and the ignition of shell O-burning. 
The common reason for these sharp increases ($^{22}$Ne) is analyzed in detail below.
At $t_{\rm cc}$\,$\simeq$\,11.5 day (seventh panel) the Si-core ignites, yielding another phase where 
$L_{\nu,{\rm nuc}}$/\Lnu\,$\simeq$\,15\%.  
At $t_{\rm cc}$\,$\simeq$\,10 hr  (last panel) the Si-shell ignites and $^{56}$Fe begins to form through $\alpha$-capture channels. Shortly after, electron capture and endothermic burning in the Fe core leads to the onset of CC. 

Overall, Figure \ref{fig:fig5} shows thermal processes are the dominant form of neutrino production until Si-depletion, when neutrinos from $\beta$-processes in Fe-group nuclei become a comparable portion of energy-loss budget until CC. In models which include more Fe-group nuclei in the nuclear network than we do here, neutrinos from $\beta$-processes surpass thermal neutrino production at the onset of CC \citep{patton_2017_aa,patton_2017_ab,farag_2020_aa}.

We calculate an approximate pair-neutrino spectrum \citep{Misiaszek_2006_p,leung_2020_aa} from
\begin{equation}
    \phi_{\rm pair}(\epsilon) = \frac{A}{k_{B}T}\left(\frac{\epsilon}{k_{B}T}\right)^{\gamma}\exp\left(\frac{-a\epsilon}{k_{B}T}\right)
    \ ,
    \label{eq:pair}
\end{equation}
where $\phi(\epsilon)$ is the number of emissions with energy $\epsilon$, and the
fitting parameters are $\alpha$\,=\,3.180657028, $a$\,=\,1.018192299, $A$\,=\,0.1425776426. This expression assumes the matter is relativistic and non-degenerate. We also assume all of the neutrinos are produced at the \Tc \ of a model, so our estimates serve as upper limits. The average pair-neutrino energy is then
\begin{equation}
    \langle\epsilon\rangle_{\rm pair} =\int_{0}^{1000} \epsilon\phi_{\rm pair}(\epsilon){\rm d}\epsilon 
    \ ,
    \label{eq:mean_pair}
\end{equation}
where the integral limits are in MeV. We also cumulatively integrate over the pair-neutrino spectrum to find the lower 10\% and upper 90\% of neutrino energies of the pair-neutrino spectrum. 

We also calculate the average electron neutrino energy $\epsilon_{\nu_e}$ from $\beta^+$ processes and average electron antineutrino energy $\epsilon_{\bar{\nu}_e}$ from $\beta^-$ processes as the sum of the energy released per second $\dot{\epsilon}_{i}$ of each weak reaction $i$  divided by the number luminosity $L_{N,i}$ 
\begin{equation}
    \epsilon_{\nu} = \sum\limits_{i=1}^N (\dot{\epsilon_{i}}) \bigg/ \sum\limits_{i=1}^N (L_{N,i})
    \ ,
    \label{eq:eq1}
\end{equation}
where $N$=40 for the low-mass reaction network and $N$=148 for the high-mass reaction network of Figure~\ref{fig:mznet}.

Figure \ref{fig:fig6} shows $\epsilon_{\nu_{e}}$ and $\epsilon_{\bar{\nu}_e}$ versus \Tc \ for different \Mzams \ at Z\,=\,1~\Zsun.  During H and He burning, $\epsilon_{\nu_{e}}$\,$\lesssim$\,1~MeV while $\epsilon_{\bar{\nu_{e}}}$\,$\simeq$\,1--1.5~MeV. From C-burning to the onset of CC, $\langle\epsilon\rangle_{\rm pair}$ remains well below $\epsilon_{\nu_{e}}$ and $\epsilon_{\bar{\nu}_e}$.  

During C and Ne burning $\beta^+$ processes are dominated by $^{21,22}$Na$\rightarrow ^{21,22}$Ne from the Ne-Na cycle, $^{26}$Al$\rightarrow ^{26}$Mg from the Mg-Al cycle, and supplemented by $^{23}$Mg$\rightarrow ^{23}$Na. These reactions decrease  $Y_e$ in the core, and produce $\nu_{e}$ with average energies $\epsilon_{\nu_e}$\,$\simeq$~1.6, 1.8, and 1.7 MeV respectively.  During this phase $\beta^-$ decays are dominated by $^{28}$Si$\leftarrow ^{28}$Al, $^{24}$Mg$\leftarrow ^{24}$Na, and $^{27}$Al$\leftarrow ^{27}$Mg, producing  $\bar{\nu}_{e}$ with average energies $\epsilon_{\bar{\nu}_e}$\,$\simeq$~1.6, 2.7, and 0.9 MeV respectively.  The total $\beta^-$ neutrino emission grows from $\simeq$\,20\% of the total $\beta$ emission during C burning to $\simeq$\,50\% during Ne burning, with $\epsilon_{\nu_{e}}$ between  1.6--2 MeV independent of \Mzams.

During Ne and O-burning there are windows where the $\epsilon_{\bar{\nu}_e}$ exceeds the $\simeq$\,1.8~MeV detection threshold to inverse beta decay of current neutrino detectors \citep[e.g.,][]{simpson_2019_aa, harada_2023_aa}. Table \ref{tab:table3} lists the dominant electron anti-neutrino luminosity sources for the $\Mzams$\,$=$\,$20 \Msun$ model during the windows where $\bar{\nu}_{e}$ exceeds current detector thresholds.

\begin{deluxetable}{lllcll}[!htb]
  \tablenum{3}
  \tablecolumns{6}
  \tablecaption{
   Potential targets for $\bar{\nu}_{e}$ detection \label{tab:table3}}
  
  \tablehead{
    \colhead{Rate} &
    \colhead{$L_{\bar{\nu}_{e}}$\!(\Lsun)} &
    \colhead{$L_{N,\bar{\nu}_{e}}$(s$^{-1}$)$^{a}$} &
    \colhead{$\epsilon_{\bar{\nu}_{e}}$(Mev)} &
    \colhead{$t_{\bar{\nu}_{e}}$$^{b}$} &
    \colhead{$t_{\bar{\nu}_{e}}$$^{b}$} 
    }
\startdata
        \code{Core Ne}\\ 
        $^{28}$Si$\leftarrow^{28}$Al & $10^{7.78}$ & 9.0$\times$10$^{46}$ & 1.6 & 7.2 d& 11.8 d\\
        $^{24}$Mg$\leftarrow ^{24}$Na & $10^{6.85}$ & 6.1$\times$10$^{45}$ & 2.7  & 5.3 d & 14 d\\
        $^{27}$Al$\leftarrow ^{27}$Mg & $10^{6.61}$ & 1.0$\times$10$^{46}$ & 0.9 & 5.2 d & 7.7 d \\
       \hline{}
       \code{Core O}\\
        $^{28}$Si$\leftarrow^{28}$Al & $10^{8.39}$ & 3.7$\times$10$^{47}$ & 1.6 & 11.7 hr & 8.3 hr \\
        $^{24}$Mg$\leftarrow ^{24}$Na & $10^{6.42}$ & 2.3$\times$10$^{45}$ & 2.8 & 13 d & 17.4 d \\
        $^{27}$Al$\leftarrow ^{27}$Mg & $10^{6.01}$ & 2.6$\times$10$^{45}$ & 1.0 & 6.5 d & 9.1 d \\     
       \hline{}
       \code{Shell O-Ne} \\
        $^{28}$Si$\leftarrow^{28}$Al & $10^{8.58}$ & 5.7$\times$10$^{47}$ & 1.6 & 13.4 hr & 19.3 hr \\
        $^{24}$Mg$\leftarrow ^{24}$Na & $10^{7.56}$ & 3.1$\times$10$^{46}$ & 1.0 & 20.4 hr & 1.24 d \\
        $^{27}$Al$\leftarrow ^{27}$Mg & $10^{7.27}$ & 4.5$\times$10$^{46}$ & 2.8 & 13.7 hr & 19.2 hr \\ 
  \enddata
  \tablenotetext{}{$^a$Anti-neutrino number luminosity.
  $^b$Time period while $L_{\bar{\nu}_{e}}$ $\geq$ 0.5 $L_{\bar{\nu}_e,{\rm max}}$. 
  $^c$Time period while $\log(L_{\bar{\nu}_{e}}/L_{\bar{\nu}_e,{\rm max}}$) $\geq$ $-$0.5.  
  All entries for the \Mzams\,=\,20~\Msun, Z\,=\,1~\Zsun \ model.}
\end{deluxetable} 

The core continues to become more neutron-rich during O-burning primarily from $^{31}$S$\rightarrow ^{31}$P, supplemented by $^{30}$P$\rightarrow ^{30}$Si, $^{36}$Ar$\rightarrow ^{36}$Cl, producing $\nu_{e}$ with
average energies $\epsilon_{\nu_e}$\,$\simeq$\, 2.2--2.4, 2.4--3.0, and 1.4 MeV respectively. $\beta$-processes in the  He, C, and Ne shells remain active. 

Core and shell Si-burning are the last exothermic burning stages and produce the Fe-peak nuclei. Initially $^{31,32}$S$\rightarrow ^{31,32}$P and $^{35,36}$Ar$\rightarrow ^{35,36}$Cl are the main $\beta$-decay channels, but are quickly replaced by $^{53,54,55}$Fe$\rightarrow ^{53,54,55}$Mn, $^{51,52,53,54}$Mn$\rightarrow ^{51,52,53,54}$Cr, $^{51,52,53,54}$Mn$\rightarrow ^{51,52,53,54}$Cr, $^{55,56,57}$Co$\rightarrow ^{55,56,57}$Fe, $^{48,49}$Cr$\rightarrow ^{48,49}$V, and $^{556,57,58,60}$Ni$\rightarrow ^{56,57,58,60}$Co.
Many of the isotopes formed during the final stages undergo $\beta$-processes that continue to make the core more neutron-rich \citep[e.g.,][]{heger_2001_aa,odrzywolek_2009_aa,patton_2017_ab} with $\epsilon_{\nu_e}$\,$\simeq$\,2.2~MeV and $\epsilon_{\bar{\nu}_e}$\,$\simeq$\,1.8~MeV.

\subsection{Six Metallicities}\label{s.hMetallicity}

\begin{figure*}[!htb]
    \centering
    \includegraphics[width=6.6in]{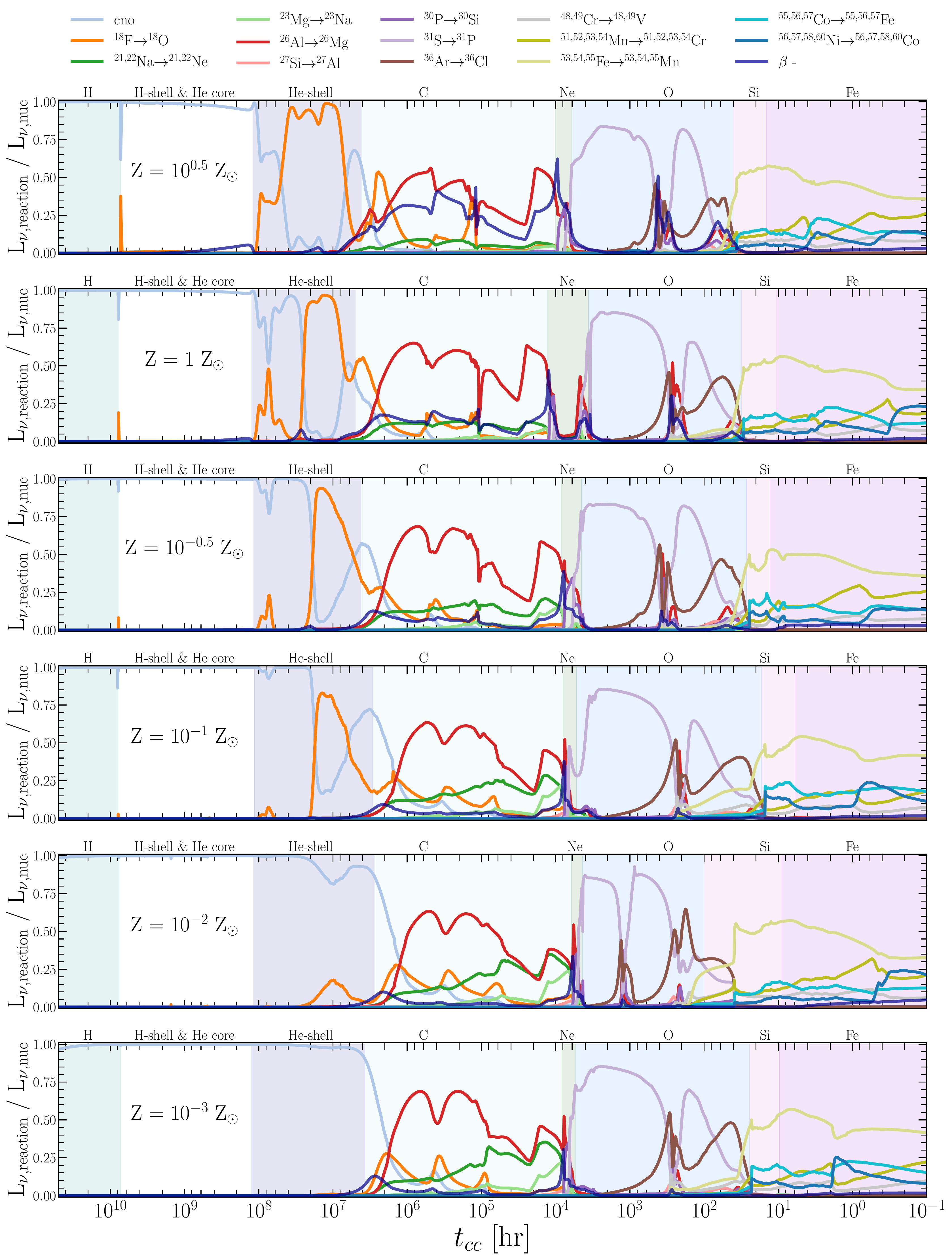}
    \caption{Components of \Lnu \ from nuclear reactions over the lifetime of a \Mzams\,=\,20 \Msun\ model for all six metallicities. The x-axis is the time to the onset of CC. Evolutionary phases are shown by the colored regions and labelled. Curves show the  largest contributions by the burning processes and weak reactions listed in the legend.}
    \label{fig:20Msun_nuclear_neu_vs_tcc}
\end{figure*}

\begin{figure*}[!htb]
    \centering
    \includegraphics[width=7.1in]{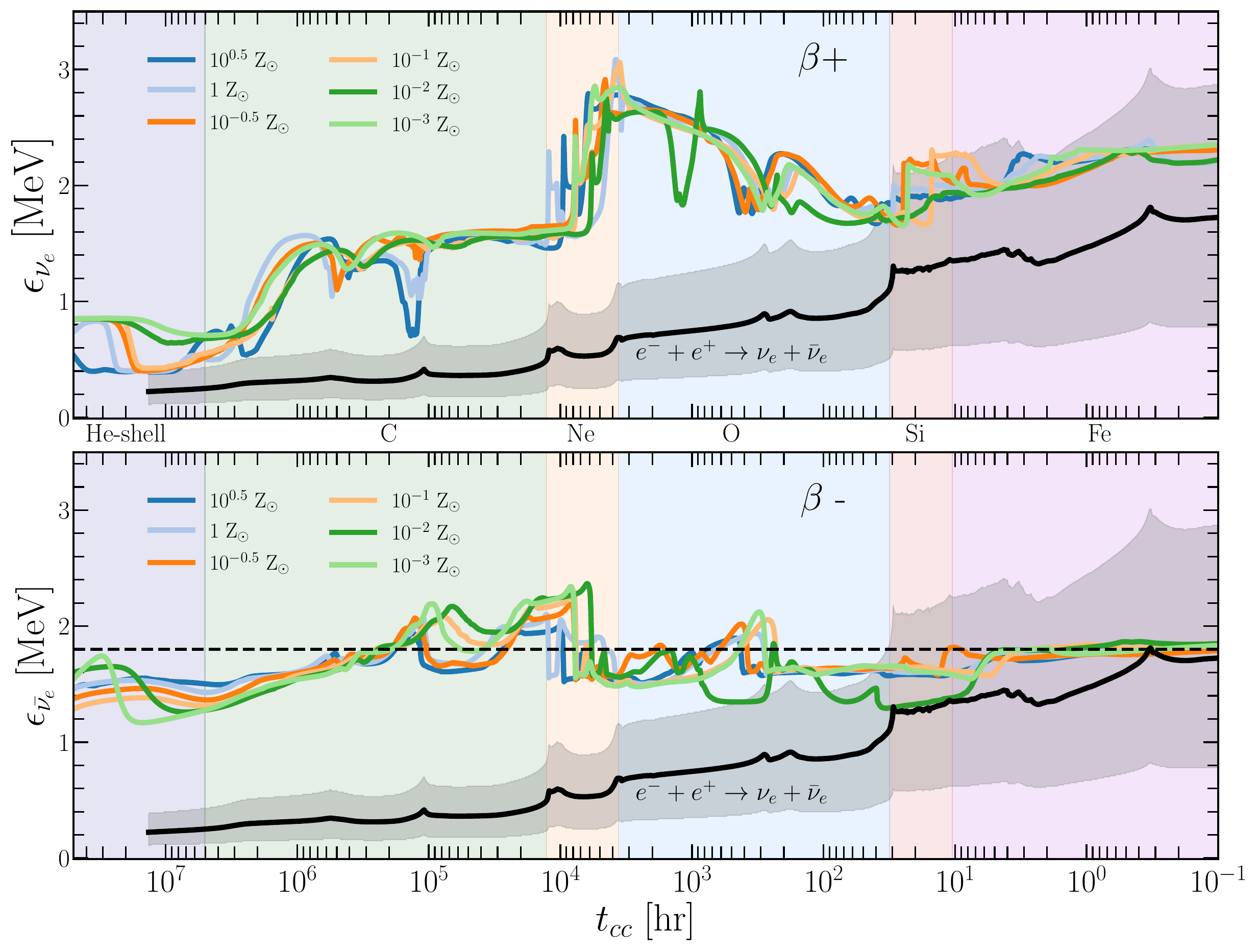}
    \caption{
    Average electron neutrino energy for beta decays (top), and average electron anti-neutrino energy for inverse-beta decays (bottom) for the \Mzams\,=\,20~\Msun \ models across all six metallicities. Curves are smoothed with a 50 timestep moving average filter. The average pair-neutrino energy is shown by the black curve, with the gray band giving the lower 10\% and upper 90\% pair-neutrino energies. Phases of evolution are shown by the colored panels and labeled. The horizontal dashed line shows a representative $\simeq$\,1.8~MeV detection threshold to inverse beta decay of current neutrino detectors \citep[e.g.,][]{simpson_2019_aa, harada_2023_aa}. The average electron neutrino and anti-neutrino energies are to first-order independent of Z. 
    }
    \label{fig:20Msun_avg_neu_energy_tcc}
\end{figure*}

Figure \ref{fig:high_mass_fighr} shows the tracks of a \Mzams\,=\,20~\Msun\ model in a $\gamma$HRD and a $\nu$HRD across all six metallicities. Overall, the low-Z models show the trend of having denser, hotter and more massive cores with lower envelope opacities, larger surface luminosities and larger effective temperatures \Teff \ than the high-Z counterparts. The hotter yet more massive H cores extends their MS lifetimes. High-Z models show significantly shorter lifetimes than low-Z models due to their smaller H abundance at the ZAMS.  For example, at the ZAMS, X\,=\,0.75 for Z\,=\,10$^{-3}$~\Zsun\ and X\,=\,0.637 for Z\,=\,10$^{0.5}$~\Zsun. The Z\,=\,10$^{0.5}$~\Zsun\ model also possesses a significantly smaller H reservoir to burn, due to the large line-driven wind mass-loss prescription ($\dot M \propto Z$) which drives the already less massive H-burning region to retreat further inward during the MS evolution, resulting in a significantly shorter MS lifetime than any other model. Metal-poor tracks have lower envelope opacities and do not evolve to as low an \Teff \ as their metal-rich counterparts. This behavior is especially prominent in the Z\,=\,10$^{-3}$~\Zsun\ model, the purple curve in Figure \ref{fig:high_mass_fighr}, which has a much shorter track in the $\gamma$HRD and is prominently offset from the lower-Z models in the $\nu$HRD. 

The first and second vertical panels in Figure~\ref{fig:20Msun_nuclear_neu_vs_tcc} show the primary source of neutrinos during H-burning and He-core burning in a 20 \Msun\ model is CNO $\beta^+$ decays. At CHeD and the onset of shell He-burning (third  vertical panel) \Lnu \ from $\beta$ decays decreases while the CO core contracts and heats up. In higher Z models, the dominant source of $\beta$ neutrinos are from $^{14}$N$\rightarrow$$^{22}$Ne in the growing He-burning shell. In lower Z models where less $^{14}$N is present, the dominant source of $\beta$ neutrinos continues to be from CNO $\beta^+$ decays in the active H-burning shell. In all models, thermally excited photoneutrinos in the hot contracting CO core begin to dominate the neutrino emission until temperatures are high enough, \Tc $\geq 7 \times 10^{8}$ K,  for pair-neutrinos to become the dominant energy loss mechanisms.

The accumulation of isotopically heavy $^{22}$Ne during He-burning provides the neutron excess $\eta$ necessary for $\beta^-$ decays to occur during advanced burning stages. A fraction of the $^{22}$Ne undergoes $^{22}$Ne($\alpha$,$n$)$^{25}$Mg and to a lesser extent $^{22}$Ne($\alpha$,$\gamma$)$^{26}$Mg. Through CHeD and into C-burning $^{22}$Ne($\alpha$,$n$)$^{25}$Mg is a neutron source for s-process nucleosynthesis  \citep{peters_1968_aa,couch_1974_aa,prantzos_1990_aa,raiteri_1991_aa,kappeler_1989_aa,gallino_1998_ab, pignatari_2010_aa,kappeler_2011_aa,wiescher_2023_aa}.

The fate of neutron-rich $^{25}$Mg evolves during C-burning \citep{raiteri_1991_ab}, which is the fusion of two $^{12}$C nuclei to form an excited $^{24}$Mg$^{*}$ nucleus which decays in three channels \citep[e.g.,][]{woosley_2002_aa}
\begin{equation} 
\begin{split}
^{12}{\rm C} + \!  ^{12}{\rm C} \rightarrow \ ^{24}{\rm Mg}^{*} & \rightarrow \  ^{20}{\rm Ne} + \alpha + \gamma \\
& \rightarrow \ ^{23}{\rm Na} + p + \gamma \\
& \rightarrow \ ^{23}{\rm Mg} + n + \gamma 
\ .
\label{eq:cburn}
\end{split}
\end{equation}
The $\alpha$- and $p$-channels occur at similar rates while the $n-$channel branching ratio of $\sim$\,1\,\% \citep{dayras_1977_aa}. Uncertainties in the branching ratios and temperature dependant rates can alter the nucleosynthetic yields during C-burning through the Ne-Na or Mg-Al cycles and the amount $^{20}$Ne available for Ne-melting \citep{bennett_2012_aa,pignatari_2013_aa,zickefoose_2018_aa,tan_2020_aa,monpribat_2022_aa}. 

The fourth vertical panel in Figure~\ref{fig:20Msun_nuclear_neu_vs_tcc} shows $^{26}$Al$\rightarrow$$^{26}$Mg (red curve) makes a primary contribution to \Lnu \ from nuclear reactions at all metallicities during C-burning.
The $p$-channel powers the Ne-Na cycle, producing a neutrino signal through $^{21,22}$Na$\rightarrow ^{21,22}$Ne and $^{23}$Mg$\rightarrow ^{23}$Na $\beta^+$ decays. $^{24}$Mg is then produced via $^{23}$Na($p$,$\gamma$)$^{24}$Mg, and $^{23}$Na($p$,$\alpha$)$^{20}$Ne creates stable $^{20}$Ne  -- now available for a later stage of Ne-melting into $^{24}$Mg and $^{28}$Si. The Mg-Al cycle is weakly powered by the $p$-channel $^{24}$Mg($p$,$\gamma$)$^{25}$Al reaction. Instead the $\alpha$-channel powers the Mg-Al cycle by providing the He nuclei necessary for $^{22}$Ne($\alpha$,$n$)$^{25}$Mg for the cycle to operate. Instead of $^{25}$Mg then being consumed by $^{25}$Mg($n$,$\gamma$)$^{26}$Mg, protons from the $p$-channel power $^{25}$Mg($p$,$\gamma$)$^{26}$Al which undergoes $\beta^+$ decay $^{26}$Al$\rightarrow ^{26}$Mg, dominating the nuclear neutrino production during C-burning.

The larger \Tc \ of low-Z models results in a stronger expression of  $^{21,22}$Na$\rightarrow ^{21,22}$Ne during C-burning. High-Z models also show a larger $\beta^-$ luminosity during C-burning than their low-Z counterparts. This results from differences in the neutron excess across metallicities. High-Z models enter C-burning with a larger $^{22}$Ne abundance available for $^{22}$Ne($\alpha$,$n$)$^{25}$Mg, which provides most of the free neutrons for an s-process \citep{raiteri_1991_ab,the_2007_aa,choplin_2018_aa}.

Another feature during C-burning is the $\beta^-$ luminosity declines from $\simeq$\,50\% of the total $\beta$ neutrino luminosity in the Z\,=\,10$^{0.5}$~\Zsun\ model, to $\simeq$\,25\% in the Z\,=\,1~\Zsun \ model, and $\leq$\,10\% in lower Z models. Independent of metallicity, these $\beta^-$ decays are primarily $^{28}$Al$\rightarrow ^{28}$Si, $^{27}$Mg$\rightarrow ^{27}$Al, and $^{24}$Na$\rightarrow ^{24}$Mg.

Neon melting is characterized by photodisintegration of Neon into $\alpha$ particles, which recapture onto a second Neon nucleus to form $^{16}$O and $^{24}$Mg.
The fifth vertical panel in Figure~\ref{fig:20Msun_nuclear_neu_vs_tcc} shows $\alpha$-capture onto the remaining $^{22}$Ne in the core provides a spike in the $\beta^-$ luminosity, and a neutron source for an s-process. A metallicity dependence on the initial $^{22}$Ne content of the core affects the strength of $\beta^-$ decays at the onset of Ne-melting. A significant fraction of the $^{26}$Mg also undergoes $^{26}$Mg($\alpha$,$n$)$^{30}$P which then decays to $^{30}$P$\rightarrow ^{30}$Si.

O-burning is the fusion of two $^{16}$O nuclei to form an excited state of $^{32}$S$^*$, which promptly decays to 
\begin{equation} \label{eq:oburn}
\begin{split}
^{16}{\rm O} + \! ^{16}{\rm O} \rightarrow \ ^{32}{\rm S}^{*} & \rightarrow \ ^{28}{\rm Si} + \alpha + \gamma \\
& \rightarrow \ ^{31}{\rm P} + p + \gamma \\
& \rightarrow \ ^{31}{\rm S} + n + \gamma \\
& \rightarrow \ ^{30}{\rm P} + d - \gamma 
\ .
\end{split}
\end{equation}
Branching ratios for the $\alpha$, $p$ $n$, and $d$ channels are $\simeq$\,34\%, 56\%, 5\%, and $\leq$\,5\% respectively, and 
the products of O-burning include $^{28}$Si, $^{32,33,34}$S, $^{35,36,37}$Cl, $^{36,37,38}$Ar, $^{39,40,41}$K, and $^{40,41,42}$Ca \citep{woosley_2002_aa}. The limited extent of neutron rich isotopes in the high-mass nuclear reaction network of  Figure~\ref{fig:mznet} means we do not capture all these isotopes, including $^{35}$S and $^{33}$P.

The sixth vertical panel in Figure~\ref{fig:20Msun_nuclear_neu_vs_tcc} shows $^{31}$S$\rightarrow ^{31}$P makes a primary contribution to \Lnu \ from nuclear reactions at all metallicities during O-burning. The accumulation of $^{36}$Ar leads to a growing neutrino signal from $^{36}$Ar$\rightarrow ^{36}$Cl. After core O-depletion, shell Ne-melting occurs before O-shell burning. The $\alpha$-captures onto the remaining $^{22}$Ne nuclei in the shell provides a second spike in the $\beta^-$ luminosity
in Figure \ref{fig:20Msun_nuclear_neu_vs_tcc}.

\begin{figure*}[!htb]
    \centering
    \includegraphics[width=6.5in]{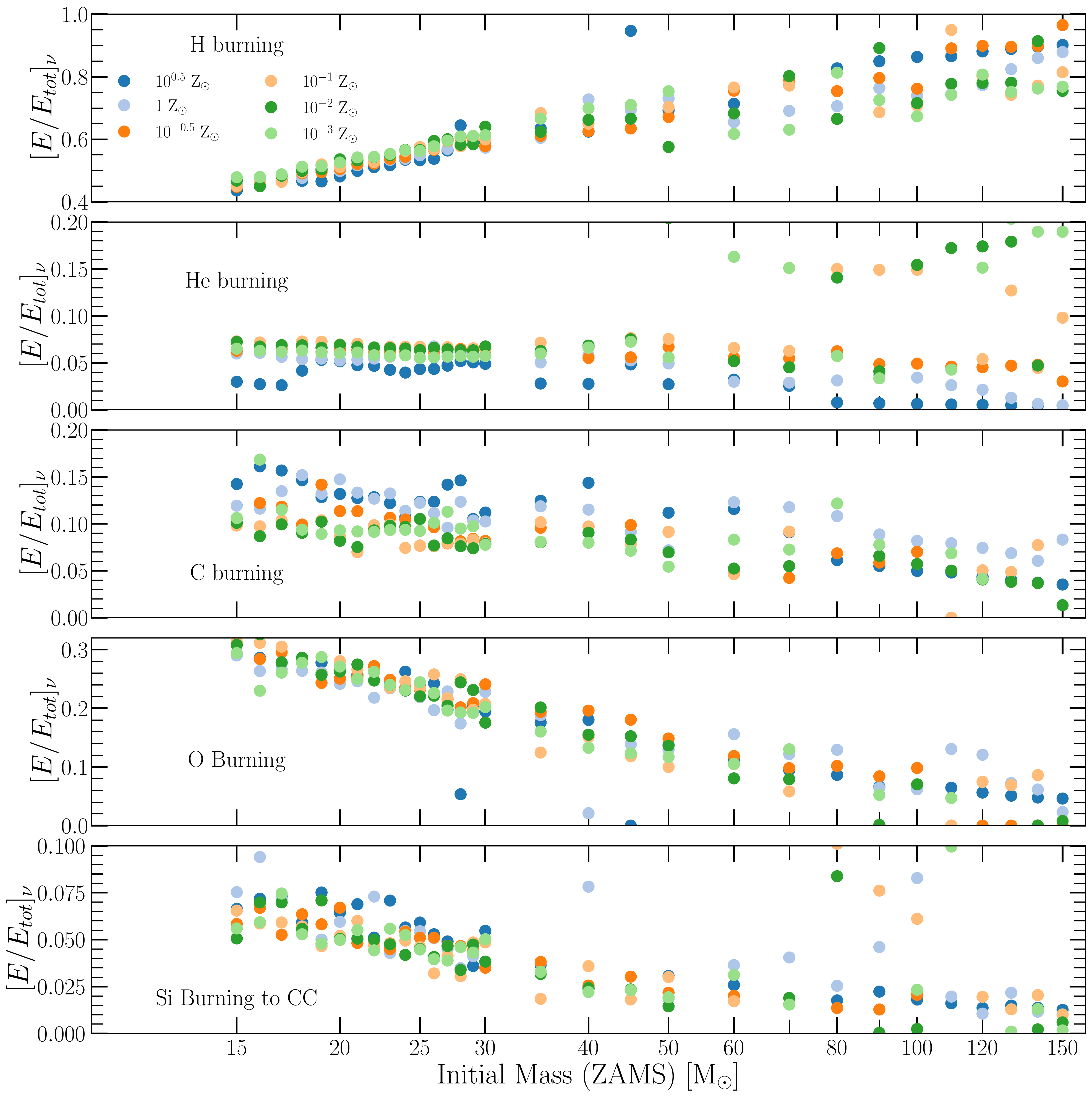}
    \caption{Fraction of \Enu\ emitted at different phases of evolution for all six metallicities (colored circles). From top to bottom, the panels show $[E/E_{\rm tot}]_{\nu}$ for H-burning, He-burning, C-burning, O-burning and Si-burning to the onset of CC.}
    \label{fig:fig20}
\end{figure*}

From Si burning (Si-$\alpha$) until CC, the seventh and eighth vertical panel in Figure~\ref{fig:20Msun_nuclear_neu_vs_tcc}, there are little differences in the relative strength of individual $\beta$ decays. At this stage of evolution, the expression of Fe-group $\beta$ decays is metallicity independent, and $\beta^-$ decays remain subdominant until $t_{cc}\lesssim 10^{-1}$ hr \citep{ patton_2017_aa, patton_2017_ab, kato_2017_aa, kato_2020_aa, kosmas_2022_aa}.

In Figure \ref {fig:20Msun_avg_neu_energy_tcc} the average neutrino and anti-neutrino energies are, to first-order, similar 
across metallicities for $\beta^+$ and $\beta^-$ decays in a 20 \Msun\ model. The largest differences in anti-neutrino energies occur during C-shell and Ne-core burning, when the neutron excess provided by $^{22}$Ne is most important. Metal-poor tracks possess lower $L_{\beta^-}$, but higher overall average anti-neutrino energy, since the signal is increasingly dominated by $^{24}$Mg$\leftarrow ^{24}$Na as opposed to $^{28}$Si$\leftarrow^{28}$Al. Windows where $\bar{\nu}_{e}$ exceeds current detector thresholds are listed in Table~\ref{tab:table3} for the Z\,=\,1~\Zsun \ model.

Figure \ref{fig:fig20} shows the fraction of the total neutrino energy produced during different phases of evolution in the mass-metallicity plane. The spread reflects the different fates experienced by stellar models of differing mass-metallicity. Larger spreads occur for  the high-mass models where wind-driven mass-loss and shell-core mergers contribute. 

Across metallicities in Figure \ref{fig:fig20}, the chief nuclear neutrino production in high-mass models come from the CNO cycle during H-burning, accounting for $\simeq$\,40--90\% of the total neutrino emission with a trend towards larger fractions with increasing \Mzams. Typical fractions for He-burning are $\lesssim 8\%$, with an exception for some very massive models that produce $\simeq$\,10--20\% from recurrent mixing of the shell-H into the He core before CHeD. Typical fractions for C-burning and O-burning are $\simeq$\,5--20\% and $\simeq 5-30$\% respectively, with a negative trend toward higher masses. From core-Si ignition to CC, $\simeq$\,2--10\% of the total neutrino emission occurs with a negative trend toward increasing masses. Overall, most neutrinos are produced during H and He burning from $\beta^+$ decays, especially in the most massive models.

\begin{figure*}[!htb]
    \centering
    \includegraphics[width=5.9in]{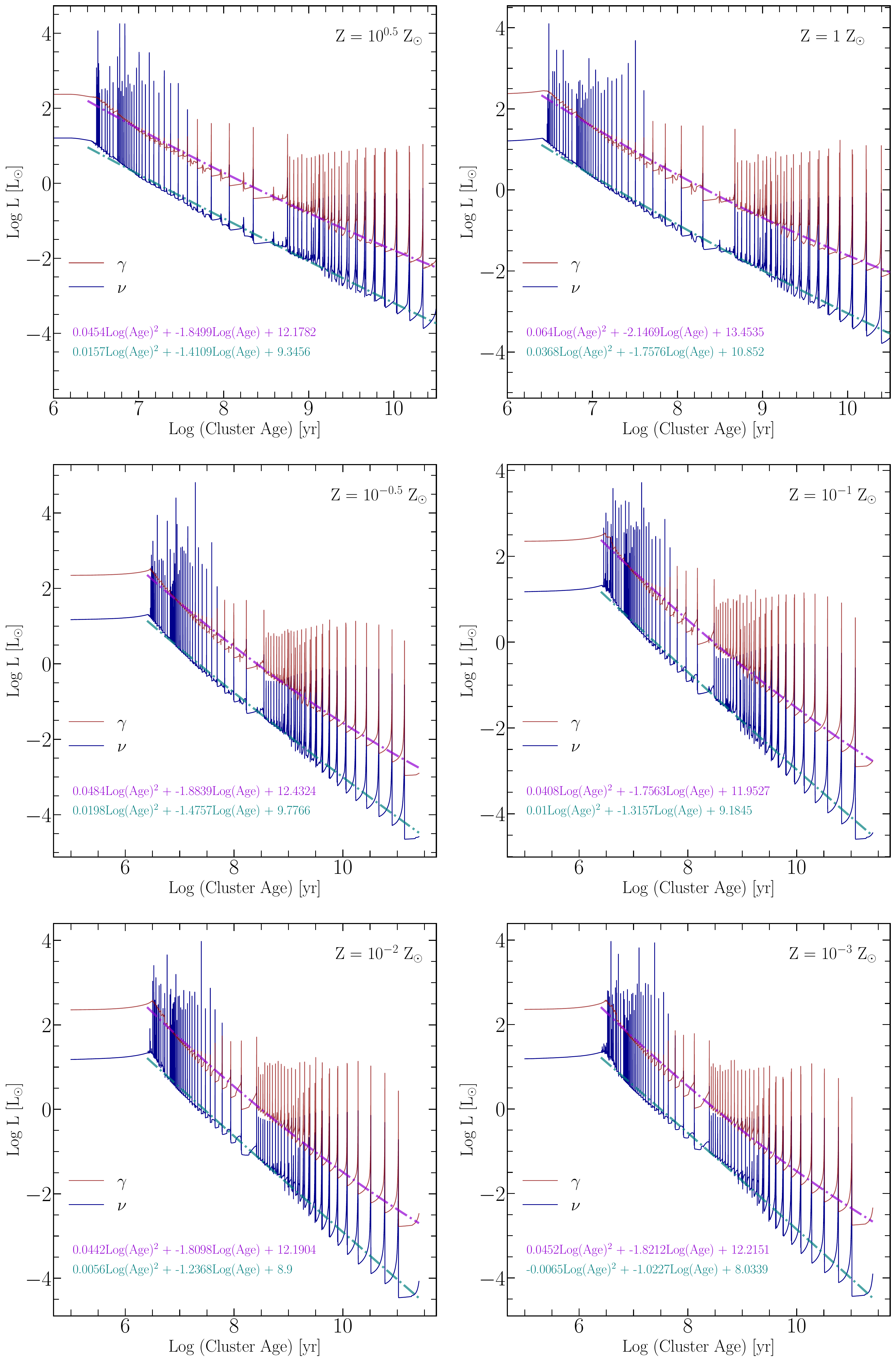}
    \caption{Cluster  \Lgamma\ and \Lnu\  light curves from the evolution models for all six metallicities. Overlayed are quadratic fitting functions with the coefficients for each metallicity listed (see Equation~\ref{eq:fit}).}
    \label{fig:fitting}
\end{figure*}

\begin{figure}[!htb]
    \centering
    \includegraphics[width=3.3in]{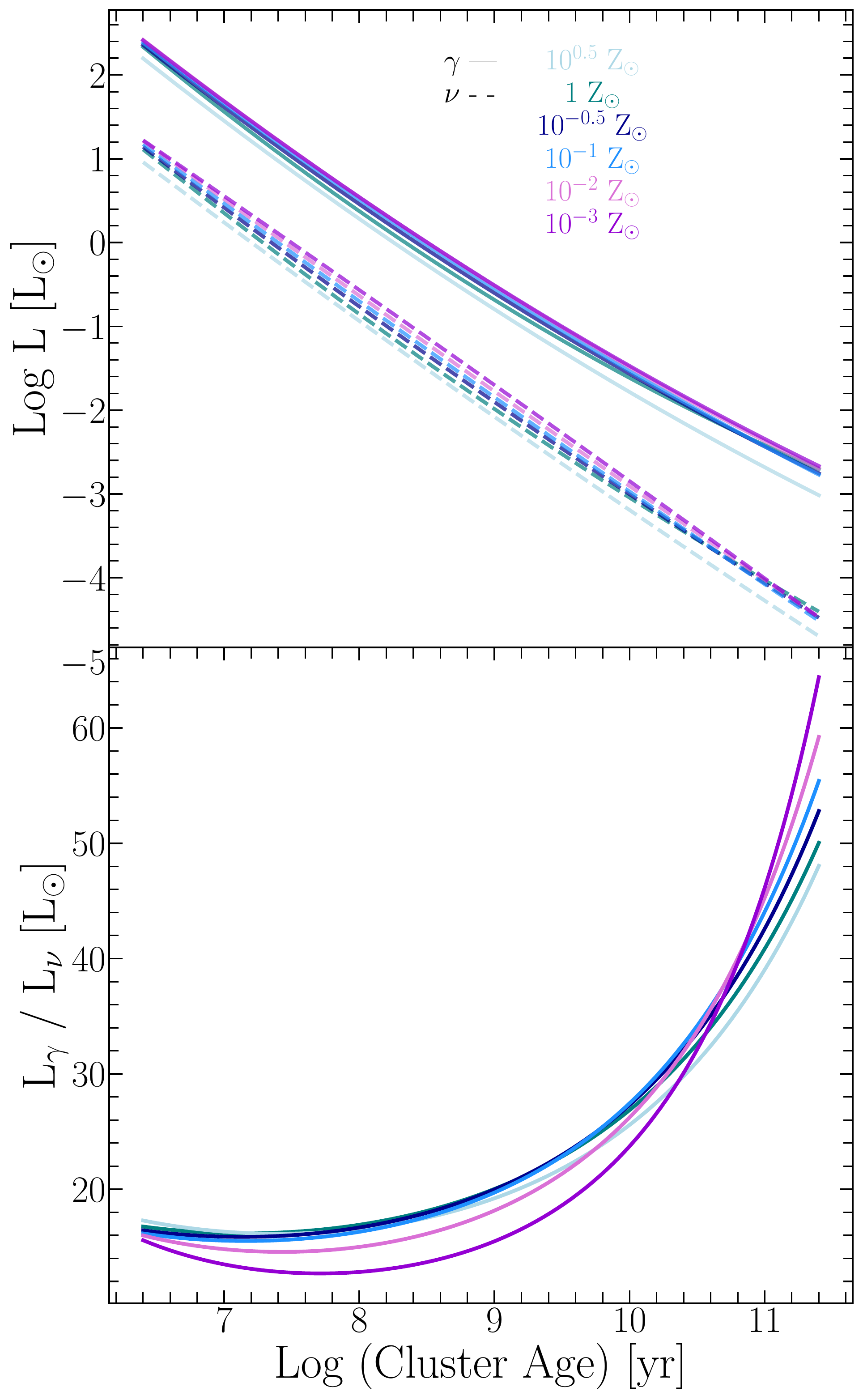}
    \caption{Cluster \Lgamma\ and \Lnu\ light curves (top) and the ratio \Lgamma/\Lnu\ (bottom) using Equation~\ref{eq:fit} with $M$\,=\,1~\Msun\ for all six metallicities.}
    \label{fig:fit}
\end{figure}

\section{Integrated Stellar Photon and Neutrino Emission}\label{s.total_production}

We explore the time-integrated photon and neutrino emission of a simple stellar population model. We assume a burst cluster population where all models are born at the same time and evolve together.
\begin{equation}
    N_{0} = \frac{1}{M_{\odot}} \int_{M_{min0}}^{M_{max}} \frac{dN}{dm} M_{*} \,dm 
    \ ,
\label{eq:eq2}
\end{equation}
\begin{equation}
    \Phi(t) = \frac{1}{N_{0}} \int_{M_{min}}^{M_{max}} \frac{dN}{dm} \phi(m,t) \,dm 
    \ ,
\label{eq:eq3}
\end{equation}
We adopt the normalized broken power law initial mass function (IMF) from \citet{kroupa_2001_aa} for the number of stars per unit mass ${dN}/{dm}$. We integrate over the IMF in Equation \ref{eq:eq2} to solve for a normalization coefficient such that a cluster of mean mass 1 \Msun\ is formed in the burst of star formation. The minimum mass $M_{\rm min0}$\,=\,0.01~\Msun\ and the maximum mass $M_{\rm max}$\,=\,150~\Msun\ of the IMF set the integration limits for the 1 \Msun\ stellar cluster. We then solve Equation \ref{eq:eq3} for $\Phi(t)$ the resultant integrated quantity, where $\phi(t)$ is the quantity we source along an isochrone. The minimum mass $M_{\rm min0}$\,=\,0.2~\Msun\ and the maximum mass $M_{\rm max}$\,=\,150~\Msun\ of the mass-metallicity plane set the integration limits.

Figure \ref{fig:fitting} shows \Lgamma\ and \Lnu\ light curves for each population synthesis model, sampled at 600,000 points in log(Age) for each metallicity. We overlay a quadratic power law for each population synthesis model to provide a convenient fitting formulae for \Lgamma \ and \Lnu \ as a function of the stellar cluster age and mass
\begin{align}
\log\left ( \frac{L}{\Lsun} \right )  & = \left [ a \log \left ( \frac{{\rm Age}}{{\rm yr}} \right )^{2} - b \log \left ( \frac{{\rm Age}}{{\rm yr}} \right ) + c \right ] \nonumber \\
 & + \log \left ( \frac{M}{\Msun} \right )
 \label{eq:fit}
\end{align}
where the fit coefficients ($a$,$b$,$c$) are listed in Table~\ref{tab:table4}.

Figure \ref{fig:fit} shows the cluster \Lgamma\ and \Lnu\ light curves and their ratio of \Lgamma/\Lnu. Both \Lgamma\ and \Lnu\ are slightly larger in low-Z models until $\sim 10^{10.5}$ Gyr when low-Z models are depleted of most H-burning and He-burning stellar tracks, and \Lgamma\ and \Lnu\ become comparable across all metallicities (except for Z = $10^{0.5}$ \Zsun).  Low-Z stellar population fits show an overall larger \Lgamma/\Lnu\ than high-Z fits until $\sim 10^{10.5}$ Gyr, when the population synthesis models are dominated by very low-mass models \Mzams\,$\leq$\,0.8~\Msun.

\begin{deluxetable}{crrr}[!htb]
  \tablenum{4}
  \tablecolumns{4}
  \tablecaption{
   Fit coefficients of Equation~\ref{eq:fit}. \label{tab:table4}}
  \tablehead{
    \colhead{ } &
    \colhead{$a$} &
    \colhead{$b$} &
    \colhead{$c$} 
    }
\startdata
        For Z\,=\,10$^{0.5}$~\Zsun\\ 
        \Lgamma &  0.0454 & 1.8499 & 12.1782 \\
        \Lnu    & 0.0157 & 1.4109 & 9.3456 \\
       \hline{}
               For Z\,=\,10$^{0.0}$~\Zsun\\ 
        \Lgamma & 0.0640 & 2.1469 & 13.4535 \\
        \Lnu    & 0.0368 & 1.7576 & 10.8520 \\
               \hline{}
               For Z\,=\,10$^{-0.5}$~\Zsun\\ 
        \Lgamma & 0.0484 & 1.8839 & 12.4324 \\
        \Lnu    & 0.0198 & 1.4757 & 9.7766 \\
               \hline{}
               For Z\,=\,10$^{-1.0}$~\Zsun\\ 
        \Lgamma & 0.0408 & 1.7563 & 11.9527 \\
        \Lnu    & 0.0100 & 1.3157 & 9.1845 \\
               \hline{}
               For Z\,=\,10$^{-2.0}$~\Zsun\\ 
        \Lgamma & 0.0442 & 1.8098 & 12.1904 \\
        \Lnu    & 0.0056 & 1.2368 & 8.9000 \\
               \hline{}
               For Z\,=\,10$^{-3.0}$~\Zsun\\ 
        \Lgamma & 0.0452 & 1.8212 & 12.2151 \\
        \Lnu    & -0.0065 & 1.0227 & 8.0339  \\
  \enddata
\end{deluxetable}
\vspace{-0.5in}

\begin{figure}[!htb]
    \centering
    \includegraphics[width=3.3in]{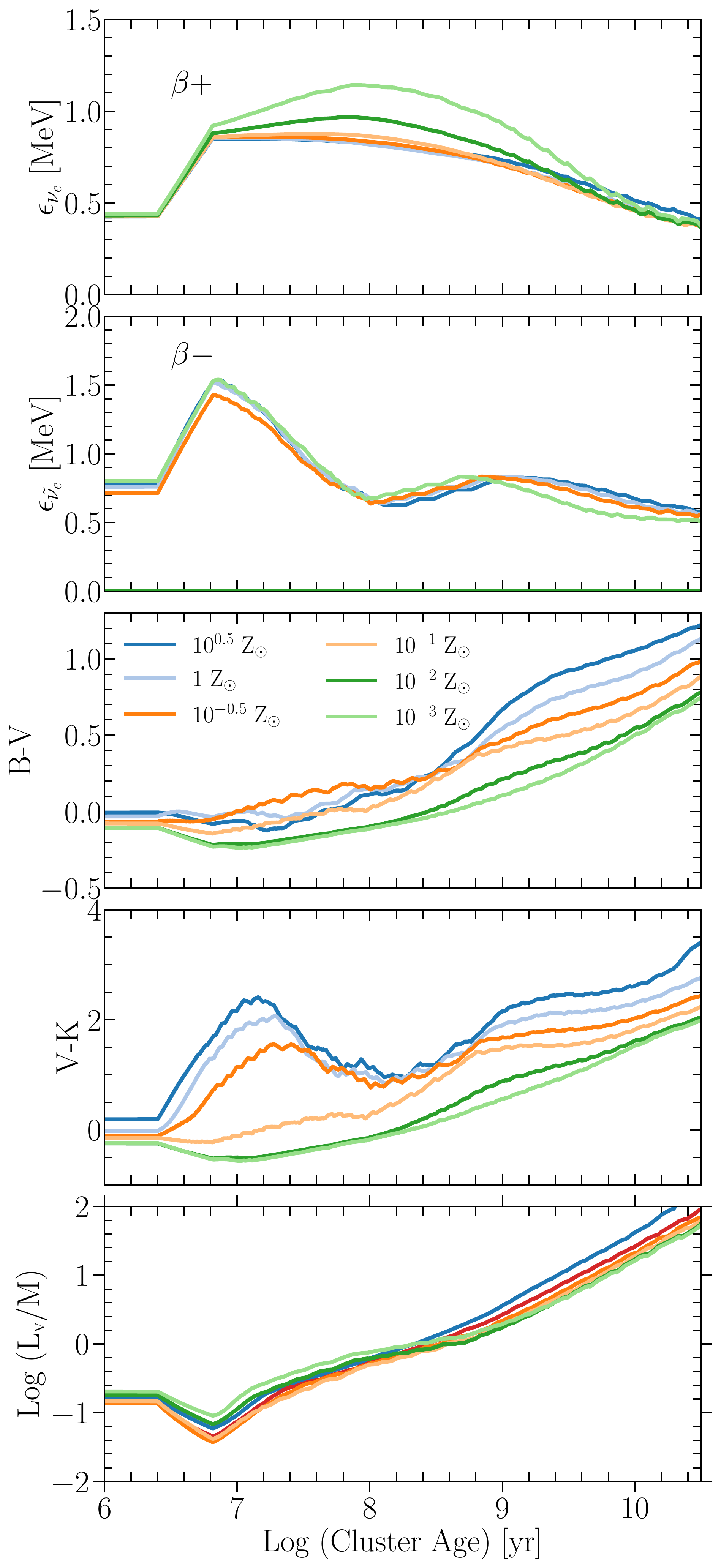}
    \caption{Average electron neutrino energies for beta decay processes (top panel), average electron anti-neutrino energies for inverse-beta decay processes (second panel), Johnson-Cousins B-V and V-K colors (third and fourth panels), and neutrino light to mass ratio for a simple stellar population (bottom panel) for all six metallicities.}
    \label{fig:fig18}
\end{figure}

Figure \ref{fig:fig18} shows the $\epsilon_{\bar{\nu}_e}$, $\epsilon_{\bar{\nu}_e}$, B-V color, V-K color, and the light to mass ratio in the V-band versus cluster age for all six metallicities. Photon and neutrino emission at early times $\simeq$\,10$^{7}$~yr is indicative of high-mass model emissions. By $\simeq$\,10$^{8}$~yr, all high-mass models have reached their final fate, leaving only low-mass models in the stellar population. Most of a star's life is spent during H and He burning in which neutrino emission is dominated by $\beta$ processes, therefore it is reasonable to approximate the average neutrino energy of a simple stellar population by $\beta$ processes alone.

The top panel in Figure~\ref{fig:fig18} shows the average neutrino energy from a simple stellar population model ranges from 0.5--1~MeV. Average neutrino energies show a slight metallicity trend, with low-Z models producing up to 0.5 MeV larger signal than high-Z models between ages of 10$^{7}$ -- 10$^{9.5}$ yr, then decreasing to $\simeq$\,0.5 MeV at 10$^{10.5}$~yr. The second panel shows the average anti-neutrino energy ranging from 0.6 -- 1.8 MeV. The anti-neutrino emission at early times, $\simeq$\,10$^{7}$~yr, is dominated by high-mass models reaching up to $\sim 1.8$ MeV. By $\simeq$\,10$^{8}$~yr, the anti-neutrino energy has reduced to $\simeq$\,0.6 MeV, and remains roughly constant until $10^{10.5}$ yr.

The third and fourth panels in Figure~\ref{fig:fig18} shows the Johnson-Cousins B-V and V-K colors respectively, calculated using the tabulations from \citet{lejeune_1998_aa}.
At early times, $\simeq$\,10$^{7}$~yr, there is a slight excess in B-V and a relatively large jump in V-K from the high-Z population models, roughly at the onset of the RSG phase in the high-mass models \citep{choi_2016_aa}. The bump in V-K is suppressed in the lowest metallicity models, which do not evolve toward the RSG branch and instead remain relatively blue, with RSG color spectra similar to the MS. At late times, $\gtrsim$\,10$^{9}$ yr when the population contains only low-mass stars, the B-V and V-K colors show an overall reddening in high-Z stellar populations. 

The V band light to mass ratio in the bottom panel of Figure~\ref{fig:fig18}  shows a weak but distinct metallicity trend. At early times, $L_{V}/M$ is larger in the lower metallicity populations. This is due to the increased \Lgamma\ in low-Z models. At late times, the trend is inverted with larger $L_{V}/M$ the high-Z population models. This is due to the longer MS lifetimes in the high-Z population models. 

\section{Summary}\label{s.conc}

We explored the evolution of stellar neutrino emission with 420 models spanning the 
initial mass 0.2\,\Msun$\leq$\,\Mzams\,$\leq$150\,\Msun\ and initial metallicity $-$3\,$\leq$\,log(Z/\Zsun)\,$\leq$\,0.5 plane.
We found lower metallicity models are more compact, hotter, and produce larger \Lnu\, with two exceptions. 
At He-core ignition on the RGB and He-shell burning on the AGB, the birth metallicity determines the amount of $^{14}$N available for the nitrogen flash $^{14}$N($\alpha$,$\gamma$)$^{18}$F(,$e^{+}\nu_e$)$^{18}$O. In high-mass models, the birth metallicity determines the amount of $^{14}$N and therefore $^{22}$Ne available for $^{22}$Ne($\alpha$,$n$)$^{25}$Mg, providing a neutron excess to power anti-neutrino emission during C, Ne and O burning. 
Overall, across the mass-metallicity plane we identify the sequence (Z$_{\rm CNO}$ $\rightarrow$ $^{14}$N $\rightarrow$ $^{22}$Ne $\rightarrow$ $^{25}$Mg $\rightarrow$ $^{26}$Al $\rightarrow$ $^{26}$Mg $\rightarrow$ $^{30}$P $\rightarrow$ $^{30}$Si) as making primary contributions to \Lnu \ at different phases of evolution.

Simple stellar populations with lower birth metallicities have higher overall \Lnu \ than their metal-rich counterparts. We find that most neutrinos from simple stellar populations are emitted in the form of electron-neutrinos through $\beta^+$ decays, with average energies in the range 0.5 -- 1.2 MeV. Lastly, we find that metal-poor stellar populations produce larger average $\beta^+$ neutrino energies (up to 0.5 MeV), though this trend is much weaker, if resolved, for $\beta^-$ neutrino emission.

We close this article by pointing out that there are many potential sensitivities that we have not investigated. Examples include choosing different convective mixing prescriptions, mass loss algorithms, and nuclear reaction rate probability distribution functions (especially $^{12}$C($\alpha$,$\gamma$)$^{16}$O and triple-$\alpha$). We also neglected rotation, their  associated magnetic fields, and binary interactions. Future uncertainty quantification studies could also explore potential couplings between simultaneous variations in uncertain parameters. We caution that these uncertainties, or missing physics, could alter the neutrino emission properties of our models.

\section*{Acknowledgements}\label{s.ack}

We thank Thomas Steindal for his helpful discussions.
We acknowledge using ChatGPT \citep{openai_2023_aa} to help polish the language of one paragraph \citep{vishniac_2023_aa}.
This research is supported by the National Science Foundation (NSF) under grant 2154339 entitled "Neutrino Emission From Stars".
This research made extensive use of the SAO/NASA Astrophysics Data System (ADS).

\facilities{Research Computing at Arizona State University.}

\software{
\MESA\ \citep[][\url{https://docs.mesastar.org}]{paxton_2011_aa,paxton_2013_aa,paxton_2015_aa,paxton_2018_aa,paxton_2019_aa,jermyn_2023_aa},
\code{MESASDK} 20190830 \citep{mesasdk_linux,mesasdk_macos},
\code{matplotlib} \citep{hunter_2007_aa}, and
\code{NumPy} \citep{der_walt_2011_aa},
\code{ChatGPT} \citep{openai_2023_aa}.
         }


\bibliographystyle{aasjournal}

\end{document}